\documentclass[11pt]{article}
\pdfoutput=1
\usepackage{jcapmod}

\usepackage{shorthand}
\usepackage{mathtools}
\usepackage{booktabs}
\usepackage[english]{babel}
\usepackage{amsmath,amssymb,amsbsy,amstext, amsthm, simplewick, amsfonts}
\usepackage{hyperref}
\usepackage{graphicx}
\usepackage[small]{caption}
\usepackage{siunitx}
\usepackage{upgreek}
\usepackage{framed}
\usepackage{wrapfig}
\usepackage{multirow}
\usepackage{bbm}
\usepackage[svgnames,dvipsnames,x11names]{xcolor}
\usepackage{tensor}

\usepackage{selinput}

\usepackage{bm}
\usepackage{float}
\usepackage{geometry}
\usepackage{yfonts}
\usepackage{caption}
\usepackage{subcaption}
\usepackage{sidecap}
\usepackage{longtable}
\usepackage{anyfontsize}
\usepackage{dsfont}
\usepackage{tikz}
\usepackage{relsize}
\usepackage{tcolorbox}

\usepackage{xcolor}
\usepackage{xparse}

\usepackage{slashed}
\usepackage{simpler-wick}

\NewDocumentCommand{\colornucleus}{omme{_^}}{%
  \begingroup\colorlet{currcolor}{.}%
  \IfValueTF{#1}
   {\textcolor[#1]{#2}}
   {\textcolor{#2}}
    {%
     #3
     \IfValueT{#4}{_{\textcolor{currcolor}{#4}}}
     \IfValueT{#5}{^{\textcolor{currcolor}{#5}}}
    }%
  \endgroup
}

\usepackage{array}
\newcolumntype{L}[1]{>{\raggedright\let\newline\\\arraybackslash\hspace{0pt}}m{#1}}
\newcolumntype{C}[1]{>{\centering\let\newline\\\arraybackslash\hspace{0pt}}m{#1}}
\newcolumntype{R}[1]{>{\raggedleft\let\newline\\\arraybackslash\hspace{0pt}}m{#1}}

\usepackage[framemethod=default]{mdframed}
\newmdenv[skipabove=7pt,
skipbelow=7pt,
rightline=false,
leftline=false,
topline=false,
bottomline=false,
backgroundcolor=gray!10,
linecolor=gray,
innerleftmargin=5pt,
innerrightmargin=5pt,
innertopmargin=5pt,
innerbottommargin=5pt,
leftmargin=0cm,
rightmargin=0cm,
linewidth=4pt]{eBox}
\newmdenv[skipabove=7pt,
skipbelow=7pt,
rightline=false,
leftline=false,
topline=false,
bottomline=false,
backgroundcolor=gray!10,
linecolor=gray,
innerleftmargin=5pt,
innerrightmargin=5pt,
innertopmargin=-5pt,
innerbottommargin=5pt,
leftmargin=0cm,
rightmargin=0cm,
linewidth=4pt]{eBox2}
\usepackage[framemethod=default]{mdframed}
\newmdenv[skipabove=7pt,
skipbelow=7pt,
rightline=true,
leftline=true,
topline=true,
bottomline=true,
backgroundcolor=gray!15,
linecolor=gray,
innerleftmargin=5pt,
innerrightmargin=5pt,
innertopmargin=5pt,
innerbottommargin=5pt,
leftmargin=0cm,
rightmargin=0cm,
linewidth=0.75pt]{eBox3}

\definecolor{Red}{RGB}{214, 39, 40}
\definecolor{Blue}{RGB} {31, 119, 180}
\definecolor{Orange}{RGB}{255, 153, 51}
\definecolor{Purple}{RGB}{178, 102, 255}
\definecolor{Green}{RGB}{44, 160, 44}

\definecolor{vio}{RGB}{19, 130, 164}
\definecolor{vioo}{RGB}{89, 2, 155}
\newcommand{\Comment}[1]{{}}
\definecolor{darkblue}{rgb}{0.15,0.35,0.55}
\definecolor{reddish}{rgb}{0.65, 0.2, 0.2}
\definecolor{darkgreen}{RGB}{50,150,0}
\definecolor{greyish}{rgb}{.90,.90,.90}
\definecolor{greyish2}{rgb}{.96,.96,.96}
\definecolor{greyish3}{rgb}{.37,.37,.37}
\definecolor{darkblue2}{rgb}{0.3,0.4,0.9}
\definecolor{Blue3}{RGB}{31, 119, 180}
\usepackage[linktocpage=true]{hyperref}
\hypersetup{
colorlinks=true,
citecolor=darkblue,
linkcolor=reddish,
urlcolor=darkblue,
pdfauthor={},
pdftitle={},
pdfsubject={}
}

\usepackage{colortbl}
\definecolor{lightgreen}{cmyk}{0.2, 0, 0.2, 0.2}
\definecolor{lightgray2}{cmyk}{0.1,0.1,0,0.1}
\definecolor{Red2}{RGB}{214, 39, 40}
\definecolor{Blue2}{RGB} {31, 119, 180}
\definecolor{Orange2}{RGB}{255, 127, 14}
\definecolor{Green2}{RGB}{44, 160, 44}

\makeatletter
\newlength{\apb@width}
\newcommand{\autoparbox}[2][c]{\settowidth{\apb@width}{#2}\parbox[#1]{\apb@width}{#2}}

\makeatother


\def\hs{\hskip 1pt}

\def\beq{\begin{equation}}
\def\eeq{\end{equation}}
\def\be{\begin{equation}}
\def\ee{\end{equation}}

\def\W{E}

\allowdisplaybreaks[1]
\begin{document}

\newgeometry{top=2cm, bottom=2cm, left=2cm, right=2cm}

\begin{titlepage}
\setcounter{page}{1} \baselineskip=15.5pt 
\thispagestyle{empty}
\vskip 28pt

\begin{center}

{\vspace{1cm}\LARGE \bf $\boldsymbol{{\rm AdS}_3}$ Recursion Relations, Double Copy\\[16pt] 
and $\boldsymbol{{\rm CFT}_2}$ Ward Identities}
\end{center}

\vskip 28pt
\begin{center}
\noindent
{\fontsize{14}{18}\selectfont 
Gr\'egoire Mathys,\hs$^{1}$
Guilherme L.~Pimentel,\hs$^{2}$  and Facundo Rost\hs$^{2}$}
\end{center}

\vspace{20pt}
\begin{center}

\vskip 12pt
\textit{$^1$ Fields and Strings Laboratory, Institute of Physics\\
Ecole Polytechnique F\'ed\'erale de Lausanne\\
CH-1015 Lausanne, Switzerland }

\vskip 12pt
\textit{$^2$ Scuola Normale Superiore and INFN\\
Piazza dei Cavalieri 7\\
56126, Pisa, Italy}
\end{center}

\vspace{0.8cm}
\begin{center}{\bf Abstract}
\end{center}
\noindent
We consider correlation functions of massless fields in three-dimensional Anti-de Sitter spacetime. Using embedding-space spinors akin to spinor helicity variables, we show that masslessness implies holomorphicity. We perform a deformation of the spinors, analogous to the BCFW shift for flat-space amplitudes, to bootstrap correlation functions of Chern--Simons theory in the bulk. We show that the resulting recursion relations in the bulk ${\rm AdS}_3$ are equivalent to the usual ${\rm CFT}_2$ Ward identities for the boundary conserved currents. The double copy of those recursion relations is equivalent to the Ward identities for the stress-energy tensor. Our findings illustrate how modern amplitude techniques can shed new light on CFT correlators, and provide a novel interpretation of fundamental results in two-dimensional conformal field theory.  
\end{titlepage}
\restoregeometry

\newpage
\setcounter{tocdepth}{3}
\setcounter{page}{2}

\linespread{1.2}
\tableofcontents
\linespread{1.1}

\newpage
\section{Introduction}
Recently, much has been learned from studying the similarities among asymptotic observables for all values of the cosmological constant: scattering amplitudes, cosmological correlators, and conformal correlation functions. This connection facilitates the interpretation of new results in each domain, and also provides inspiration for analogous structures that exist in one situation, but might not yet exist in another. The constraints of symmetry and masslessness are particularly strong, and oftentimes corner us into a small menu of possibilities. This is most transparent in the right kinematical variables. 

\vskip 4pt

This paper is about a particular instance of this simplicity: masslessness in three-dimensional Anti-de Sitter (AdS) space is made manifest by using spinor variables, which we describe in detail and that are quite similar to those used in four-dimensional flat space. In these variables, we write recursion relations for bulk correlation functions in a similar fashion to that for tree-level scattering amplitudes. The resulting relations have a beautiful interpretation on the two-dimensional boundary: they are the celebrated Ward identities of the current (or stress-energy tensor) in a Conformal Field Theory. Our main contribution is to make this connection transparent by using a suitable kinematic language.

\vskip 4pt

In a little more detail: in \cite{Baumann:2024ttn} we emphasized that masslessness in spacetimes with a cosmological constant is tied to the notion of holomorphicity in the ``right" variables. In four-dimensional curved space, that required the introduction of twistors. In this paper we apply the same idea to a simpler setup, that of three-dimensional Anti-de Sitter space (or, alternatively, of two-dimensional Conformal Field Theory). In this arena, boundary positions are represented as projective null rays $P^M$, with $P^2=0$, on the light-cone of four-dimensional embedding space. They can then be parameterized in terms of embedding-space spinors, in the same way as spinor helicity variables parameterize on-shell massless four-momenta $p^\mu$, with $p^2=0$, in four-dimensional flat space. The resulting kinematic constraints are slightly different, but much of the analysis of boundary correlators proceeds by analogy as in flat space. In particular, imposing masslessness (in the bulk) requires correlation functions to be holomorphic in the spinor variables.\footnote{Similar spinor techniques have been proposed for Conformal Field Theories, see e.g.~\cite{Maldacena:2011nz,Caron-Huot:2021kjy,Jain:2021vrv,Skvortsov:2022wzo}.} In two-dimensional CFTs this is the usual notion of holomorphicity for currents and tensors that makes their conservation manifest. For higher-dimensional CFTs the analogous notion is more involved and is most transparent in twistor space. For recent progress in this direction, see~\cite{Bala:2025gmz,CarrilloGonzalez:2025qjk,Bala:2025jbh,Bala:2025qxr,Ansari:2025fvi,Arundine:2026fbr,De:2026shn,Bala:2026hdm,Bala:2026bdx,Huang:2026tsh,Arundine:2026myr,Bala:2026trw,CarrilloGonzalez:2026eum,Bala:2026hag,Arundine:2026qqg}.

\vskip 4pt

From the constraints of holomorphicity, we explore the correlation functions of massless gauge fields and (boundary) gravitons in three-dimensional Anti-de Sitter space. Our main results include a reinterpretation of the Belavin-Polyakov-Zamolodchikov (BPZ) Ward identities for conserved currents and the stress-energy tensor \cite{BELAVIN1984333}. From a ``bulk" perspective, they follow from tree-level recursion relations---{\it \`a la} Britto-Cachazo-Feng-Witten (BCFW)---for (boundary) gluons and  gravitons.\footnote{In three dimensions, the graviton has no propagating degrees of freedom, while the gluon has as leading bulk interaction in three dimensions a Chern--Simons term, which enforces that the local curvature always vanishes.} Moreover, we find a double-copy prescription of gluon correlators at any number of points that gives graviton correlators. Ultimately, this is intimately related to the well-known Sugawara construction of the stress-energy tensor for CFTs with a Kac-Moody symmetry.

\vspace{0.25cm}
\paragraph{Outline} In Section 2, we briefly review the embedding-space formalism for CFTs, introduce spinor variables in embedding space for two-dimensional CFTs, and demonstrate their utility in bootstrapping two- and three-point correlation functions. In Section 3, we further use this machinery to derive recursion relations for conserved currents and show how they reproduce Ward identities in Chern--Simons theories for AdS$_3$. In Section 4, we reproduce this for the stress-energy tensor and reinterpret the double copy in this formalism. Section 5 contains our conclusions and outlook. Two appendices collect technical details.

\paragraph{Notation}

Two-dimensional positions are denoted by $x^\mu$, with $\mu=1,2$, while the holomorphic and antiholomorphic coordinates are $w=x^1+ix^2$ and $\bar w=x^1-ix^2$, respectively. The corresponding positions in embedding space are $P^M$, with $M=0,1,2,3$ for $\mathbb{R}^{1,3}$, and we use $\W^M$ for the polarization vectors in embedding space. For two-dimensional CFTs, we define spinor variables for the embedding-space position $P^M$ in the same way as spinor helicity variables in four-dimensional flat space:
\be 
P_{\alpha\dot\alpha}\equiv P_M\hs(\sigma^M)_{\alpha\dot\alpha}=\Lambda_\alpha\tilde \Lambda_{\dot\alpha}\,.
\ee 
Here, $\Lambda_\alpha$ and $\tilde \Lambda_{\dot \alpha}$ are two-component spinors, and we contract $P_M$ with the Pauli matrices $(\sigma^M)_\alpha^{~\dot\beta}=(\delta_\alpha^{~\dot\beta},(\sigma^i)_\alpha^{~\dot\beta})$ with one index lowered by the Levi-Civita symbol $\epsilon_{\dot\alpha\dot\beta}$ (with $\epsilon^{12}=\epsilon_{21}=1$) following the left-multiplication convention
\be 
v_\alpha=\epsilon_{\alpha\beta }v^\beta \,,\qquad\quad  v^\alpha=\epsilon^{\alpha\beta }v_\beta\,.
\ee 
We also define angle and square spinor brackets as
\be
\langle ij \rangle \equiv \epsilon^{\dot\beta\dot\alpha}\tilde\Lambda_{i,\dot\alpha} \tilde\Lambda_{j,\dot\beta}\,, \qquad \quad [ij] \equiv \epsilon^{\beta\alpha}\Lambda_{i,\alpha} \Lambda_{j,\beta} \,,
\ee
respectively.

 \section{Spinors for Embedding Space}

In this section, we review the embedding space as a means of trivializing the constraints imposed by conformal symmetry. We write the coordinates of embedding space using a pair of spinors, in an analogous way to how it is done for flat space scattering amplitudes. Then we study the consequences of conformal symmetry in embedding-space spinors, focusing on conserved operators. We emphasize that holomorphicity in these spinor variables trivializes current conservation for correlators.

\subsection{Review of Embedding Space}

We will begin with a lightning review of the embedding-space approach to conformal field theories. Originally, embedding space was introduced as a particularly powerful way to make conformal symmetry manifest. This comes at the price of obscuring how to impose current conservation for (massless) spinning fields. This is meant as a brief introduction for novices, and experts should feel free to skip it. More details can be found in~\cite{Rychkov:2016iqz}.

\subsubsection*{Projective null cone} 

Conformal transformations are nonlinear transformations and they become particularly involved for spinning operators. Nevertheless, it is somewhat straightforward to show that the conformal algebra on $d$-dimensional Euclidean space $\mathbb{R}^d$ is isomorphic to the algebra of Lorentz transformations on $(d+2)$-dimensional Minkowski space $\mathbb{R}^{1,d+1}$. This suggests that it should be possible to find a suitable embedding of $\mathbb{R}^d$
into $\mathbb{R}^{1,d+1}$ such that conformal transformations
become as simple as Lorentz transformations (i.e. linear) on the lower-dimensional slice. This is the goal of the so-called embedding-space formalism of conformal field theory, which goes back to
Dirac~\cite{Dirac:1936fq}. Its recent resurgence has been central to many new developments in the CFT literature, e.g.~\cite{Costa:2011mg,Costa:2011dw,Karateev:2017jgd}.

\vskip 4pt
The basic idea is the embedding of $\mathbb{R}^d$ as a slice through a higher-dimensional light-cone. In particular, consider the $(d+2)$-dimensional Minkowski spacetime $\mathbb{R}^{1,d+1}$ with coordinates
\be 
 P^M, \qquad \text{with}\qquad  M=0,1,\cdots,d+1\, ,\label{eq:Embeddingcoord}
\ee  and metric $\eta^{MN}=\text{diag}(-1,1,\cdots,1)$.  Under a Lorentz transformation, these spacetime coordinates transform as $P^M \mapsto L^M{}_N P^N$ with $L^M{}_N$ an $SO(1,d+1)$ matrix, and the goal is to find an embedding where these become conformal transformations.  The first step is to consider solely points living on the 
{\it projective null cone} in the embedding space, which means 
\begin{align}
	P^2 &=0\,,\label{eq:p2vanish}\\
	P^M &\sim \rho P^M\,, \label{equ:rescaling}
\end{align}
where $\rho \in \mathbb R_{>0}$ is a rescaling parameter.
The condition \eqref{eq:p2vanish} is Lorentz invariant and serves to remove one of the coordinates defined in \eqref{eq:Embeddingcoord} in terms of all the other ones. To remove a second coordinate such that we obtain a $d$-dimensional spacetime, we define a \textit{section} of the light-cone $P^+\equiv f(P^\mu)$, where $P^{\pm} = P^{0}\pm P^{d+1}$ are light-cone coordinates and the coordinates $P^\mu$ with $\mu=1,\dots,d$ are identified with the coordinates $x^\mu$ on $\mathbb{R}^d$.

\vskip 4pt

\begin{figure}[t!]
	\centering
\includegraphics[scale=0.6]{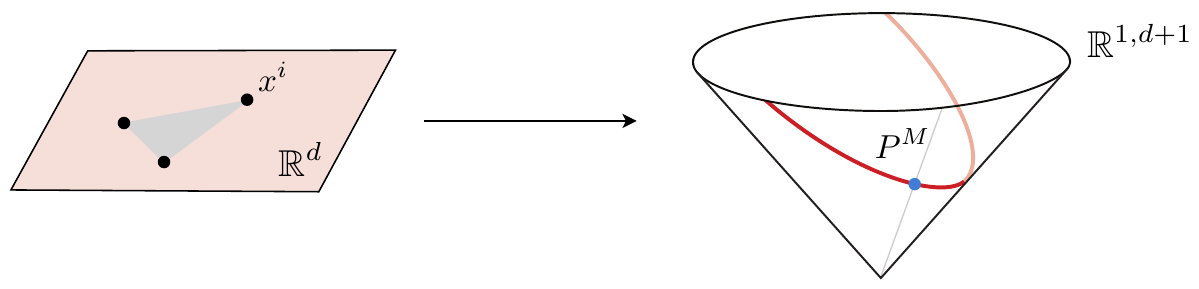} 
	\caption{In the embedding-space formalism, we map points on $\mathbb{R}^d$ to a section of a projective null cone in~$\mathbb{R}^{1,d+1}$. Lorentz transformations in the ambient space become conformal transformations on this section.}
    \label{fig:Embedding}
\end{figure}

For concreteness, we choose $f(P^\mu)=1$. It can then be shown (see for example \cite{Rychkov:2016iqz}) that Lorentz transformations on the embedding space, when combined with the projective equivalence~\eqref{equ:rescaling} imply conformal transformations on the {\it Euclidean section}
\be
(P^+,P^-, P^\mu) = (1,x^2,x^\mu)\, .
\label{equ:section}
\ee
In Figure~\ref{fig:Embedding}, we illustrate  the action of a Lorentz transformation on an infinitesimal interval~${\rm d} x$. At first, it looks like this transformation moves the interval off the Euclidean section, but, under the identification (\ref{equ:rescaling}), we get the new interval ${\rm d} x^\prime$ on the section. It can also be shown that the constraint (\ref{equ:section}) is designed in such a way that the combined Lorentz transformation and rescaling becomes a conformal transformation on the Euclidean slice~\cite{Rychkov:2016iqz}.

\subsubsection*{Tensors in embedding space}
To be able to describe spinning fields, we need to understand how to handle these in embedding space. For concreteness, we consider a symmetric, traceless and transverse tensor $O_{M_1 \ldots M_S}$ defined on the projective null cone $P^2=0$. Under the rescaling \eqref{equ:rescaling}, the tensor transforms as 
\be 
O_{M_1 \ldots M_S}(\rho P) = \rho^{-\Delta}O_{M_1 \ldots M_S}(P)\, ,\label{eq:ScalingTensor}
\ee 
which implies that it is a homogeneous function of degree $-\Delta$. In fact, it is convenient to contract the tensor components with auxiliary null polarization vectors $\W^M\in \mathbb{R}^{1,d+1}$, such that the tensors can be written in index-free notation, 
\be
O^{(S)}(P,\W) = O_{M_1 \ldots M_S}(P) \,\W^{M_1} \cdots \W^{M_S}\,.\label{eq:tensortoscalar}
\ee
The polarization vector $\W^M$ satisfies $\W^2=P\cdot \W=0$. Note that this introduces an extra ``gauge invariance" under $\W\rightarrow\W + c P$, which together with the scaling \eqref{eq:ScalingTensor} exactly removes two components per index from the tensor in embedding space such that its independent components match with those of the tensor on the Euclidean section.  Under
rescalings of both the embedding-space coordinates and the polarization vectors, we have
\beq
O^{(S)}(\rho P, \alpha \W) = \rho^{-\Delta} \alpha^S \,O^{(S)}(P,\W)\, ,
\label{equ:scaling}
\eeq
where $\Delta$ and $S$ are the scaling dimension and spin of the field $O^{(S)}$, respectively.

\vskip 4pt

This machinery implies that correlators in the $d$-dimensional Euclidean space are lifted to homogeneous functions on the light-cone of the $(d+2)$-dimensional Minkowski spacetime, where the conformal group acts as the Lorentz group. 

\subsubsection*{Conformal correlators}

In embedding space, conformal correlators are simply built from the most general Lorentz-invariant expressions with the correct scaling behavior according to \eqref{equ:scaling}. To make progress, it is convenient to define the (parity-even) conformally-invariant structures 
\beq
\begin{aligned}
	P_{ij} &\equiv -P_i \cdot P_j\,,\\[6pt]
	H_{ij} &\equiv - 2 \big[(\W_i \cdot \W_j)(P_i \cdot P_j) - (\W_i \cdot P_j)(\W_j \cdot P_i)\big]\,,\\
	V_{i,jk} &\equiv \frac{(\W_i \cdot P_j)(P_k \cdot P_i) - (\W_i \cdot P_k)(P_j \cdot P_i)}{P_j \cdot P_k}\,,
\end{aligned}
\label{equ:BuildingBlocks}
\eeq
which serve as the basic building blocks for conformal correlators. As a simple example, the two-point function of field $O^{(S)}$ of spin-$S$ and conformal weight $\Delta$ is forced to take the form
\be
\braket{O_1^{(S)} O_2^{(S)}} \sim \frac{H_{12}^S}{(P_{12})^{\Delta+S}} \, ,
\label{equ:2ptEmb}
\ee
where we have dropped an overall normalization constant. The subscript on $O_i^{(S)}$ denotes both the type of the field and its position, i.e.~$O_i^{(S)} = O_i^{(S)}(P_i^M)$. The right-hand side of equation~\eqref{equ:2ptEmb}
is the only combination that can be constructed using $P_{12}$ and $H_{12}$ that has the appropriate rescaling covariance \eqref{equ:scaling} for each field. Similarly, the three-point function of two scalars and a
spin-$S$ field is
\be
\braket{O_1 O_2 O_3^{(S)}} \sim \frac{V_3^S}{P_{12}^{(\Delta_1+\Delta_2-\Delta_3-S)/2} P_{23}^{(\Delta_2+\Delta_3-\Delta_1+S)/2} P_{31}^{(\Delta_3+\Delta_1-\Delta_2+S)/2} } \,,
\ee
where we defined $V_i \equiv V_{i,jk}$ for $\{i,j,k\}$ a cyclic permutation of $\{1,2,3\}$, and dropped an overall constant factor.

\vskip 4pt
There are also examples where more than one structure is consistent with conformal invariance. This is the case, for instance, for three identical spin-$2$ operators which have 5 allowed structures. Generically, the number of allowed structures grows as we increase the spin of the fields \cite{Costa:2011mg}.

\subsubsection*{Current conservation} Correlators of conserved tensors must satisfy an additional kinematic constraint, which in position space reads\footnote{Note that this holds as an operator equation, but when inserted inside correlators, this only holds up to contact terms.}
\beq
\partial_{\mu_n}J^{\mu_1 \ldots \mu_n \ldots \mu_S}=0\,.
\eeq
As long as the scaling dimension of the conserved current takes the value 
$\Delta=S+d-2$,
this conservation condition can be uplifted to embedding space as~\cite{Costa:2011mg} 
\be 
\frac{\partial}{\partial P_{M_n}}D_{M_n} J=0\,,\quad\text{where}\quad D_M \equiv \left(\frac{d-2}{2}+\W\cdot \frac{\partial}{\partial \W}\right)\frac{\partial}{\partial \W^M}-\frac{1}{2}\W_M\frac{\partial^2}{\partial \W\cdot \partial \W}\,.
\label{equ:conservation-operator-embedding}
\ee  
To impose conservation for a certain correlator, we then deduce
the linear combinations of the conformally-invariant structures  that are compatible with the conservation condition.

\vskip 4pt

Conceptually, it is somewhat unsatisfactory that we first have to list the larger space of conformally-invariant structures and then find the specific linear combination(s) that are compatible with the constraint \eqref{equ:conservation-operator-embedding}. It would be much more satisfactory if we could directly write down the structures where both conformal invariance and conservation are manifest from the start. To achieve this goal, we will define spinors in embedding space, where a notion of holomorphicity will trivialize both constraints simultaneously.

\subsection{Spinor Variables in Embedding Space\label{subsec:spinorhelicity}}

Spinor helicity variables are a particularly useful way to describe the on-shell kinematics of scattering amplitudes of massless particles in four dimensions; see \cite{Elvang:2015rqa, Cheung:2017pzi} for a review. The starting point is the observation that
on-shell massless particles satisfy $p^2=0$, such that their four-momenta can be written as  
\beq
p_{\alpha \dot \alpha} = p_\mu (\sigma^\mu)_{\alpha \dot \alpha}  = \lambda_\alpha \tilde\lambda_{\dot{\alpha}}\,.
\label{equ:4d-spinors}
\eeq
Here, we contract $p_\mu$ with the Pauli matrices $(\sigma^\mu)_\alpha^{~\dot\beta}=(\delta_\alpha^{~\dot\beta},(\sigma^i)_\alpha^{~\dot\beta})$ including the identity, with one index lowered by~$\epsilon_{\dot\alpha\dot\beta}$ (i.e.~$(\sigma^\mu)_{\alpha\dot\alpha}=\epsilon_{\dot\alpha\dot\beta}(\sigma^\mu)_\alpha^{~\dot\beta}$), and $\lambda_\alpha$ and $\tilde\lambda_{\dot \alpha}$ are two-component spinors.  Given two particles $i$ and $j$, their ``angle" and ``square" brackets are defined as 
\be\label{equ:4d-spinor-brackets}
\braket{ij} \equiv\epsilon^{ \dot{\beta}\dot{\alpha}  } \tilde{\lambda}_{i,\dot{\alpha}} \tilde{\lambda}_{j, \dot{\beta}}\hs  \,, \qquad [ij] \equiv \epsilon^{   \beta  \alpha}\lambda_{i, \alpha}  \lambda_{j, \beta}\,,
\ee
where $\epsilon^{\alpha \beta}$ is the Levi-Civita symbol (with $\epsilon^{12}=+1$).  These brackets are the Lorentz-invariant building blocks of the spinor helicity formalism. This implies that any invariant function of the four-dimensional on-shell kinematics can be written in terms of these brackets. For example, the familiar Mandelstam variables are
$s_{ij} = -(p_i + p_j )^2 = -2p_i \cdot p_j = -\braket{ij} [ij]$.

\vskip 4pt

The main benefit of these variables is to make the symmetries of the problem manifest, and Lorentz invariance is guaranteed as long as the on-shell amplitudes are written in terms of these brackets. Moreover, by inspection of (\ref{equ:4d-spinors}), we see that the momentum $p^\mu$ is left invariant if the two spinors are multiplied by inverse complex factors
\beq\label{eq-littlegroup-flat}
\lambda \mapsto t^{-1} \lambda\,,\quad \tilde \lambda \mapsto t \tilde \lambda \, ,
\eeq
where $t$ is either a phase $e^{i\theta}$ or a complex number for real or complex momenta, respectively. 
The U(1) symmetry \eqref{eq-littlegroup-flat} is called the {\it little group}. Covariance under these little group transformations is an important element in the bootstrap of on-shell scattering amplitudes; see \cite{Elvang:2015rqa, Cheung:2017pzi} for a detailed discussion.

\subsubsection*{Spinors for embedding space} 
We explained that the coordinates of the embedding space $P^M\in \mathbb{R}^{1,d+1}$ (as defined in \eqref{eq:Embeddingcoord}) satisfy
 $P^2=0$ (see \eqref{eq:p2vanish}). The fact that $p^2=0$ is the crucial observation that renders spinor helicity variables so useful for massless amplitudes. It  thus seems natural to wonder whether these variables can also help streamline calculations in embedding space. In the following, we will show that this is indeed the case.

\vskip 4pt
 
 For two-dimensional CFTs, this is particularly simple as their embedding space is simply four-dimensional Minkowski space, which is the natural habitat of the usual spinor helicity variables. There are nevertheless two important differences compared to the use of this formalism for flat-space amplitudes: 
\begin{enumerate}
\item Unlike momentum space four-vectors, the embedding coordinates do {\it not} satisfy the analog of momentum conservation, i.e. $P_1+P_2 + \cdots + P_N \neq 0$.
\item The little group that leaves the physical position $x^\mu$ invariant is slightly larger than the little group in 4d momentum space. This is because a rescaling of the spinors that corresponds to the rescaling $P^M \rightarrow \rho P^M$ does not change the corresponding physical position $x^\mu$ on the Euclidean slice. This implies that these rescalings also belong to the little group. 
\end{enumerate}
 As will become clear,  these two differences compensate each other and allow spinor helicity variables to efficiently describe CFT correlators in embedding space.
 
 \vskip 4pt
 Mimicking (\ref{equ:4d-spinors}), we write the embedding-space position $P^M \in \mathbb{R}^{1,3}$
  in terms of spinor variables 
\be\label{eqdefspinors}
P_{\alpha\dot{\alpha}}\equiv P_{M}(\sigma^M)_{\alpha\dot{\alpha}}=\Lambda_\alpha \tilde\Lambda_{\dot{\alpha}}\,,
\ee  
where $\Lambda_\alpha$ and $ \tilde\Lambda_{\dot \alpha}$ are two-component spinors, and $(\sigma^M)_{\alpha\dot\alpha}$ are the Pauli matrices with one index lowered as defined below~\eqref{equ:4d-spinors}. Similarly to \eqref{equ:4d-spinor-brackets}, we define spinor brackets which will serve as the basic building blocks for conformal correlators as
\be
\langle ij \rangle \equiv \epsilon^{\dot\beta\dot\alpha}\tilde\Lambda_{i,\dot\alpha} \tilde\Lambda_{j,\dot\beta}\,, \qquad [ij] \equiv \epsilon^{\beta\alpha}\Lambda_{i,\alpha} \Lambda_{j,\beta} \,.
\ee
The distance between two points $P_i$ and $P_j$ is
\be
(P_i-P_j)^2=-2P_i\cdot P_j =\braket{ij}[ij]\,.
\ee
Notice that the transformations
\be
\tilde\Lambda \mapsto r\, \tilde\Lambda\,,\qquad 
\Lambda \mapsto  \bar r \,\Lambda\,,
\ee 
change the embedding-space position as $P^M\mapsto r \bar r P^M$, where $r$ and $\bar r$ may be arbitrary complex parameters if we complexify the embedding-space positions, or the spinors $\Lambda$ and $\tilde \Lambda$. Since $P^M \sim r\bar r P^M$, this leaves the position
$x^\mu$ invariant and thus these transformations belong to the little group. Unlike in the 4d flat-space case~\eqref{equ:4d-spinors}, the little group now has two independent parameters. Indeed, writing $r= \rho t$ and $\bar r=\rho/t$, we see that the little group contains the usual transformation $\tilde\Lambda\mapsto t \hs\tilde\Lambda$ and $ \Lambda\mapsto t^{-1}\Lambda$ that comprises the little group of 4d flat space (see equation \eqref{eq-littlegroup-flat}). Moreover, it also contains another transformation $\Lambda\mapsto \rho \hs\Lambda$ and $\tilde \Lambda\mapsto \rho\hs\tilde\Lambda$, which amounts to a rescaling of the spinor variables that in turn yields a rescaling $P^M\mapsto \rho^2P^M$ of the embedding-space position. This implies that the angle and square brackets transform as
\be
\label{equ:bracket-trans}
\begin{aligned}
\braket{ij} &\mapsto r_i r_j \braket{ij} 
\,,\\
[ij] &\mapsto \bar r_i  \bar r_j  [ij] \, .
\end{aligned}
\ee
These transformations will be important for bootstrapping conformal field theory correlators below.

\vskip 4pt
Like the polarization vectors in 4d spinor helicity variables, 
the auxiliary vectors $\W^M$ in (\ref{eq:tensortoscalar}) can be expressed in terms of reference spinors $\eta$ and $\tilde\eta$:
\begin{align}
	\W_{\alpha\dot{\alpha}}^+&\equiv \W_{M}^+\sigma^M_{\alpha\dot{\alpha}}=\sqrt{2}~\displaystyle\frac{\Lambda_{\alpha}\tilde\eta_{\dot\alpha}  }{\braket{\tilde\eta \tilde\Lambda}}\,,\\
	\W_{\alpha\dot{\alpha}}^-&\equiv \W_{M}^-\sigma^M_{\alpha\dot{\alpha}}=\sqrt{2}~\displaystyle\frac{\eta_{\alpha}\tilde\Lambda_{\dot\alpha}  }{[\eta \Lambda]}\,,
\end{align}
which explicitly satisfy $\W\cdot P=\W^2=0$. Notice that $\tilde\eta_{\dot\alpha}\mapsto \tilde\eta_{\dot\alpha}+c\tilde\Lambda_{\dot\alpha}$ amounts to a gauge transformation $\W^+\mapsto \W^++c P$, which leaves the embedding-space tensor invariant, $\Phi(P,\W^+ + c P)  = \Phi(P,\W^+)$. A similar transformation applies to $\eta_{\alpha}$ and $\W^-$. It is thus clear that correlators cannot depend on the choice of the reference spinors, as they are invariant under this gauge transformation.

\vskip 4pt
Under the little group, the auxiliary vectors transform as
\be
\begin{aligned}
\W^+ &\mapsto t^{-2}\hs \W^+ \,,\\
\W^- &\mapsto t^2\hs \W^- \,.
\end{aligned}
\ee
Combining this with the scaling behavior in (\ref{equ:scaling}), we find how a tensor, with fixed polarization, transforms under the little group:
\begin{align}
\Phi(P,\W^+)&\mapsto \Phi\left(\rho^2 P,t^{-2}\hs\W^+\right)=\rho^{-2\Delta} t^{-2S} \Phi(P,\W^+)\,, \\
\Phi(P,\W^-)&\mapsto \Phi\left(\rho^2 P,t^{+2}\hs\W^-\right)=\rho^{-2\Delta} t^{+2S}  \Phi(P,\W^-)\,.
\end{align}
Defining $\Phi^{(\Delta,s)} = \Phi^{(\Delta,\pm S)}\equiv \Phi(P,\W^\pm)$, we encode the choice of polarization in the sign of the so-called ``planar spin" $s\equiv \pm S$. This allows us to write the little group transformation for the field as
\be 
\Phi^{(\Delta,s)}\mapsto \rho^{-2\Delta} t^{-2 s} \Phi^{(\Delta,s)}=
 r^{-2 h}\bar{r}^{-2 \bar h}\hs \Phi^{(\Delta,s)}\,,
\label{equ:LG}
\ee
where we have defined $h \equiv (\Delta+s)/2$ and $\bar h \equiv (\Delta-s)/2$ (which we will later identify as the conformal weights of the field). To avoid clutter, we will henceforth drop the superscript on~$\Phi^{(\Delta,s)}$.
A generic $N$-point correlator then transforms under the little group as 
\be\label{eqcorrelatorlittlegroup}
\braket{\Phi_1\cdots \Phi_N} \mapsto \left(\prod_{i=1}^N \rho_i^{-2\Delta_i} t_i^{-2s_i}\right) \braket{\Phi_1\cdots \Phi_N} \equiv \left(\prod_{i=1}^{N} r_i^{-2h_i} \bar{r}_i^{-2\bar{h}_i}\right)\braket{\Phi_1\cdots \Phi_N}\,,
\ee
which is the same as for helicity amplitudes in flat space, except for the additional rescaling under dilatations. 

\vskip 4pt

Let us clarify what the symmetry and little group generators are in our embedding-space treatment. Consider the following differential operators
\be\label{equ:def-operator-F}
F^{\dot{\alpha}\dot{\beta}}\equiv \tilde{\Lambda}^{\dot{\alpha}} \frac{d}{d\tilde{\Lambda}_{\dot{\beta}}}\,,\quad
\bar{F}^{\alpha\beta} \equiv  \Lambda^\alpha \frac{d}{d \Lambda_\beta}\,.
\ee
The symmetric parts of these operators are the generators of global conformal transformations:
\be\label{eq-cft2-generators}
J^{\dot{\alpha}\dot{\beta}} \equiv F^{(\dot{\alpha}\dot{\beta})}\,,\quad
\bar{J}^{\alpha\beta} \equiv \bar{F}^{(\alpha\beta)}\,.
\ee
It is easy to check that these satisfy $\mathfrak{sl}(2)\oplus \hs\mathfrak{sl}(2)$ commutation relations, which is the complexification of the $\mathfrak{so}(1,3)$ Lie algebra of global conformal transformations. The antisymmetric parts of these operators, on the other hand, are the generators of little group transformations:
\be
\begin{aligned}
	H&\equiv -\frac{1}{2} \epsilon_{\dot{\beta}\dot{\alpha}}F^{\dot{\alpha}\dot{\beta}}= -\frac{1}{2}\tilde\Lambda_{\dot{\beta}} \frac{d}{d\tilde\Lambda_{\dot{\beta}}}\,,\\
	\bar{H}&\equiv -\frac{1}{2} \epsilon_{\beta\alpha}\bar{F}^{\alpha\beta}=-\frac{1}{2}  \Lambda_\beta \frac{d}{d\Lambda_\beta}\,.
\end{aligned}
\ee
Due to the little group transformation law (\ref{equ:LG}), the correlators are homogeneous functions of the spinors $\tilde\Lambda$ and $\Lambda$ with degree $-2h$ and $-2\bar{h}$, respectively. The eigenvalues of the little group generators $H$ and $\bar{H}$ are therefore $h$ and $\bar{h}$, respectively. Moreover, we can define the generator of scalings as $D\equiv H+\bar{H}$, with eigenvalue $h+\bar{h}=\Delta$, and the generator of U(1) rotations as $L\equiv H-\bar{H}$, with eigenvalue $h-\bar{h}=s$.

\subsubsection*{Position space} To make contact with the standard literature on two-dimensional CFT, it is convenient to study what this construction represents in ordinary position space. We start by projecting
 the above spinors to the physical positions on the Poincar\'e slice via
\be\label{eqpoincareslice}
\tilde\Lambda_{\dot\alpha}=(1,w)\,,\qquad\Lambda_{\alpha}=(-\bar w,1)\,,
\ee
where $w$ and $\bar{w}$ are complex numbers, which will be conjugates of each other when imposing the reality condition. The embedding-space position that corresponds to this choice of spinors is 
\be 
P^M=-\frac{1}{2}(\sigma^M)_{\alpha\dot{\alpha}}\hs \Lambda^{{\alpha}}\tilde \Lambda^{\dot\alpha}=(P^+,P^-,P^1,P^2)=\left(1,w\bar{w},\frac{w+\bar{w}}{2},\frac{w-\bar{w}}{2i}\right) ,
\ee
which can be identified with the coordinates on the Poincar\'e slice $P^M =(1,\vec{x}^{\hs 2},x^1,x^2)$ in equation~\eqref{equ:section} provided 
\be
w \equiv x^1+ix^2\,,\qquad 
\bar{w} \equiv x^1-i x^2\,.
\ee
This implies that the angle and square brackets on the Poincar\'e slice are
\be\label{eqspinorbraketpoincareslice}
\begin{aligned}
\langle ij\rangle &=w_i-w_j
\,,\\
[ij] &=\bar{w}_i-\bar{w}_j\, .
\end{aligned}
\ee
In other words, the angle and square brackets simply describe the holomorphic and antiholomorphic separations between two points in position space.

\subsection{Bootstrapping Spinning Correlators} 
\label{sec:bootstrap}
Now that we have gone through the somewhat lengthy process of defining these spinor variables, it is time to illustrate why they are useful. As a first application, we will now use these variables in embedding space to ``bootstrap" the two- and three-point functions of spinning fields. 

\subsubsection*{Two-point functions}
To determine the two-point functions, we start with the most general ansatz
\be 
\braket{\Phi_1\Phi_2}=\frac{\text{const.}}{\langle 12 \rangle^{n} [12]^{\bar n}}\,,
\label{equ:2pt}
\ee
where conformal invariance is simply reflected in the fact that the result is solely written in terms of spinor brackets. We are left with determining the exponents, for which we will use the scaling under little group transformations. In particular, spinor brackets transform as in (\ref{equ:bracket-trans}), while the correlator transforms as in (\ref{eqcorrelatorlittlegroup}). This implies that under a little group transformation, the two sides of (\ref{equ:2pt}) transform as
\be 
\begin{aligned}
\label{eq:Mapping2pt}
\braket{\Phi_1\Phi_2} &\mapsto r_1^{-2h_1} r_2^{-2h_2} \bar r_1^{-2\bar h_1}\bar r_2^{-2\bar h_2} \braket{\Phi_1\Phi_2}\,,\\[4pt]
\frac{1}{\braket{12}^{n} [12]^{\bar n}} & \mapsto r_1^{-n} r_2^{-n} \bar r_1^{-\bar n}\bar r_2^{-\bar n}  \frac{1}{\braket{12}^{n} [12]^{\bar n}}\, . 
\end{aligned}
\ee
This is consistent only in the case where  
\be
\begin{aligned}
n &= 2h_1 = 2h_2 \equiv 2h\,,\\
\bar n&= 2\bar h_1 =2\bar h_2 \equiv 2\bar h\, ,
\end{aligned}
\ee
such that the two-point function is completely fixed to be 
\be 
\braket{\Phi_1\Phi_2}=\frac{\text{const.}}{\langle 12 \rangle^{2h} [12]^{2\bar h}}\, .
\label{equ:2pt-2}
\ee
This is indeed the two-point function of a primary field of conformal weight $(h,\bar h)$ and planar spin $s=h-\bar h$~\cite{Ginsparg:1988ui,DiFrancesco:1997nk}. Note that in deriving \eqref{equ:2pt-2}, we never forced the fields $\Phi_i$ to have identical weights, but this came out directly as a consequence of imposing little group covariance. In the usual treatment of CFT correlation functions, this requirement is a consequence of imposing invariance of the correlator under special conformal transformations.

\subsubsection*{Three-point functions} For the three-point function, 
the starting ansatz is 
\be\label{equ:3pt}
\braket{\Phi_1 \Phi_2 \Phi_3}=\frac{\text{const.}}{\braket{12}^{n_3}[12]^{\bar{n}_3}\braket{23}^{n_1}[23]^{\bar{n}_1}\braket{31}^{n_2}[31]^{\bar{n}_2}}\, ,
\ee
where we have chosen the labeling of the exponents for later convenience. Notice that this ansatz is similar to that for massless three-particle amplitudes in flat space. There is an important difference though: for amplitudes, the ansatz involves only angle (square) brackets in the holomorphic (antiholomorphic) configuration. This is due to the fact that momentum conservation $p_1+p_2+p_3=0$ implies that either all angle brackets or all square brackets are vanishing. On the other hand, in these spinor variables for embedding space, we clearly do not have a condition like $P_1+P_2+P_3=0$, and thus we allow the ansatz to have both types of spinor brackets.
\vskip 4pt

However, we have a useful tool at our disposal that was not available in flat space. In particular, we can use the homogeneous scaling of the fields, which is given by  $\Phi(\rho P,\W)=\rho^{-2\Delta} \Phi(P,\W)$. This arises because, besides the spatial rotation $SO(2)$, there is yet an extra generator of the little group that was not there in flat space, namely that of dilatations. Thus, the contracted tensors $\Phi(P,\W)$ and the correlators should be covariant under dilatations as well as under spatial rotations. Indeed, the correlators must transform under the full little group as \eqref{eqcorrelatorlittlegroup}, and imposing this transformation law for the ansatz in equation \eqref{equ:3pt}, it is straightforward to show that the exponents of the ansatz must be
\begin{align}
n_k &=h_i + h_j - h_k\,,\\
\bar{n}_k &= \bar h_i + \bar h_j - \bar h_k\,,
\end{align}
where $\{i,j,k\}$ is a cyclic permutation of $\{1,2,3\}$. As before, $h_i \equiv (\Delta_i +s_i)/2$ and $\bar{h}_i \equiv (\Delta_i-s_i)/2$ are the conformal weights of the field $i$.

\vskip 4pt

Interestingly, although the symmetry group is smaller for CFT$_2$ correlators than for scattering amplitudes in four dimensions due to the absence of translation invariance, the little group is bigger. This compensates for the lack of translation symmetry, and allows us to bootstrap three-point correlators in terms of these spinor variables in embedding space in a very similar way to the bootstrapping of three-point amplitudes in spinor helicity variables.

\subsubsection*{Conserved tensors} 
In a unitary 2d CFT, a conserved, traceless (quasi-)primary current of spin $|s|$ saturates the unitarity bound $\Delta\geq |s|$, so each of its independent components has weights $(h,\bar h) = (|s|,0)$ or $(0,|s|)$ according to the sign of its spin $s=h-\bar h$. The mixed components, which would have $h,\bar h$ both nonzero, vanish by tracelessness. The current therefore decomposes into a holomorphic component $T(w)$ (with $\bar h=0$) and an antiholomorphic component $\bar T(\bar w)$ (with $h=0$), which are separately conserved, i.e. they satisfy $\bar{\partial}T = \partial\bar T=0$. Thus, the two- and three-point correlators of these conserved currents with generic spins are given by 
\begin{align}
\braket{T_1T_2}& =\frac{\text{const.}}{\langle 12 \rangle^{2|s|} }\,, \quad \braket{T_1 T_2 T_3 } =\frac{\text{const.}}{\braket{12}^{|s_1|+|s_2|-|s_3|}\braket{23}^{|s_2|+|s_3|-|s_1|}\braket{31}^{|s_3|+|s_1|-|s_2|}}\,,\\
\braket{\bar T_1\bar T_2}& =\frac{\text{const.}}{[12 ]^{2|s|} }\,, \quad \braket{\bar{T}_1\bar{T}_2\bar{T}_3} =\frac{\text{const.}}{[12]^{|s_1|+|s_2|-|s_3|}[23]^{|s_2|+|s_3|-|s_1|}[31]^{|s_3|+|s_1|-|s_2|}}\,.
\end{align}
These correlators are either holomorphic or antiholomorphic in the sense that they only depend on either angle or square brackets, respectively. After projecting to the Poincar\'e slice via \eqref{eqspinorbraketpoincareslice}, the two cases become dependence on $
w_i$ alone or on $\bar{w}_i$
 alone, so the correlators are holomorphic or antiholomorphic in physical position space. In 2d CFTs conservation is famously tied to holomorphicity, and here that link takes a particularly simple form: taking the correlators to be purely holomorphic (or purely antiholomorphic) makes conservation manifest.

\vskip 4pt

Moreover, notice that holomorphic correlators correspond to correlators with all-plus polarizations, meaning that all of the fields have polarization vectors $\W_i^+$, i.e.~all of them satisfy $s_i>0$. Analogously, antiholomorphic correlators correspond to correlators with all-minus polarizations $\W_i^-$, i.e. all fields satisfy $s_i<0$.

 \section{Correlators of Spin-1 Currents}

As usual in conformal field theory, the two- and three-point functions were fixed completely by symmetries, and our spinor variables in embedding space have given us an efficient way to bootstrap the answers.
To determine higher-point functions, we need to supply additional dynamical information. In this section, we will show that these higher-point functions can be constructed through an analog of the BCFW recursion relations~\cite{Britto:2004ap,Britto:2005fq} by using the OPE between our fields as the input coming from dynamics. Ultimately, these recursion relations are equivalent to the well-known BPZ Ward identities. 

\subsection{Recursion Relations}

Borrowing inspiration from the recursion relations of scattering amplitudes that we briefly describe in Appendix \ref{sec:amplitudes-recursion} for non-experts, we will first study a complex deformation of the spinor variables in embedding space that probes the (complexified) kinematics of the boundary correlators without referring to the bulk. We will then use this together with the OPE of conserved currents in spinor space to derive a recursion relation for correlators of holomorphic conserved currents. Using recursion relations in AdS correlators has a long history dating back to \cite{Raju:2010by,Raju:2012zr} (see also, e.g., ~\cite{Zhou:2018sfz,Armstrong:2022mfr} for more recent developments). The deformation we use here is similar in spirit, with the advantage of preserving all the bulk isometries.

\subsubsection*{Complex deformations}
Let us consider the following (complex) deformation of the embedding-space positions:
\be
P_i^M\mapsto P_i^M(z)= P_i^M+z \hs Q_i^M\, ,
\ee
for some vector in embedding space $Q_i^M$. Since there is no analog of flat-space momentum conservation $\sum_i p_i^\mu=0$ in embedding space, these embedding-space positions do not need to satisfy $\sum_i P_i^M=0$. This is an important difference compared to massless scattering amplitudes in 4d flat space. However, the analog of the constraint $p_i^2=0$, of course, still holds in embedding space because the deformed coordinates must still lie on the null cone, $P_i^2(z)=0$ for any $z$. The shift vectors $Q_i^M$ must therefore satisfy
\be
Q_i^2=Q_i\cdot P_i=0\,.
\label{equ:constraints}
\ee
Writing the embedding-space positions as $2\times 2$ matrices $P_{i;\alpha\dot{\alpha}}=P_{i,M} (\sigma^M)_{\alpha\dot{\alpha}}$, the constraint~\eqref{equ:constraints} together with $P_i^2=0$ is equivalent to $\det(P_{i;\alpha \dot{\alpha}}(z))=0$ for any value of $z$. 
It is straightforward to show that any deformation that satisfies this constraint is of one of the two following forms:
\begin{align}
	P_{i;\alpha \dot{\alpha}}(z) &=\Lambda_{i,\alpha}\tilde{\Lambda}_{i,\dot\alpha}(z)\,,\quad\text{with}\quad\tilde{\Lambda}_{i,\dot\alpha}(z)= \tilde{\Lambda}_{i,\dot{\alpha}} + z\hs \tilde{\eta}_{i,\dot{\alpha}}\,,\\  P_{i;\alpha \dot{\alpha}}(z)&= \Lambda_{i,\alpha}(z) \hs\tilde\Lambda_{i,\dot{\alpha}} \,,\quad\text{with}\quad \Lambda_{i,\alpha}(z)=\Lambda_{i,\alpha}+ z \hs\eta_{i,\alpha}
	\, .
\end{align}
In other words, this implies that we only deform the embedding-space spinor $\tilde{\Lambda}_{i,\dot{\alpha}}$ or $\Lambda_{i,\alpha}$ defined in Section \ref{subsec:spinorhelicity}. The specific choice of deformation is defined by the spinor $\tilde\eta$ (or $\eta$), which is linearly independent of $\tilde\Lambda_i$, so that $\langle i\tilde\eta\rangle\neq 0$.

\vskip 4pt
Take, for instance, an $n$-point correlator and deform only the tilded spinor  $\tilde\Lambda_1$ corresponding to the first field as
\begin{equation}
\tilde\Lambda_{1,\dot{\alpha}}\mapsto \tilde\Lambda_{1,\dot{\alpha}}(z)=\tilde\Lambda_{1,\dot\alpha}+z\tilde{\eta}_{\dot\alpha}\,,
\end{equation}
while leaving all the other spinors fixed. In terms of brackets, this implies that we deform only the following
angle brackets
\be
\braket{i1}\mapsto \braket{i1(z)}=\braket{i1}+z\braket{i\tilde\eta}\,, \label{equ:bracket}
\ee
for $i=2,\, 3,\, \dots,\,n$, while leaving all the remaining angle and square brackets invariant. Notice that the deformed spinor $\tilde\Lambda_1(z)$ scales in the same way as the undeformed spinor $\tilde\Lambda_1$, as long as the parameter $z$ transforms as $z\mapsto r_1\hs r_{\tilde\eta}^{-1} \hs z$ under the rescalings $\tilde\Lambda_1\mapsto r_1\hs\tilde\Lambda_1$ and $\tilde\eta\mapsto r_{\tilde\eta}\hs\tilde\eta$. 

\vskip 4pt

For explicit computations it is convenient to work on the  Poincar\'e slice. We use~\eqref{eqpoincareslice}  for the \textit{undeformed} spinor $\tilde\Lambda_{1,\dot{\alpha}}^P\equiv(1,w_1)$. The reference spinor can likewise be put on the slice, $\tilde\eta_{\dot{\alpha}}^P\equiv(1,w_{\tilde\eta})$, at the cost of a rescaling that we absorb into a redefinition of $z$. The result of deforming the spinor is then
\be\label{eqdeformedspinorpoincare}
\tilde\Lambda_{1,\dot \alpha}(z)=\tilde\Lambda^P_{1,\dot \alpha}+z\tilde\eta^P_{\dot \alpha}=(1+z,w_1+zw_{\tilde\eta})=(1+z)\left(1,\frac{w_1+zw_{\tilde\eta}}{1+z}\right) . 
\ee
An overall rescaling then puts the deformed spinor $\tilde\Lambda_1(z)$ onto the Poincar\'e slice
\be
\tilde\Lambda^P_{1,\dot \alpha}(z)\equiv (1+z)^{-1} \tilde\Lambda_{1,\dot \alpha}(z)=\left(1,\frac{w_1+zw_{\tilde\eta}}{1+z}\right)\equiv (1,w_1(z)) \,, 
\ee
so that  the deformed holomorphic position is
\be
w_1(z)=\frac{w_1+zw_{\tilde\eta}}{1+z}\,.
\label{equ:Def-w1}
\ee
Of course, the remaining holomorphic positions $w_i$ (for $i\neq 1$), and all the antiholomorphic positions $\bar{w}_j$ are left invariant, because we are not deforming $\tilde\Lambda_i$ (for $i\neq 1$) nor $\Lambda_j$.

\subsubsection*{Singularities}
We can now study the complex shift  $ \tilde\Lambda_1(z)=\tilde\Lambda_1+z\tilde\eta$ to the $n$-point function of holomorphic conserved currents:
\be\label{eq-deformed-correlator}
\braket{JJ\cdots J}(z)=\braket{J(\tilde\Lambda_1(z))J(\tilde\Lambda_2)\cdots J(\tilde\Lambda_n)}\,,
\ee
where we have suppressed the color indices $a_i$ of the currents $J_i$.
Note that this complex deformation probes the complexified kinematics where the embedding-space positions or the spinor variables can take any complex values, and are not restricted by any reality condition. Thus, there is an implicit step here given by analytically continuing the correlator to complex embedding-space positions or spinor variables.

\vskip 4pt

Assuming that the deformed correlator is a meromorphic function of the deformation parameter $z$, the undeformed correlator can then be written as a sum of residues (including possibly the residue at $\infty$):
\begin{equation}\label{eq-residue-sum}
	\braket{JJ\cdots J}(z=0) =\oint \frac{dz}{2\pi i} \frac{\braket{JJ\cdots J}(z)}{z}
	=-\sum_I \underset{z=z_I}{\rm Res}\left(\frac{\braket{JJ\cdots J}(z)}{z}\right) .
\end{equation}
To relate the correlator written in terms of the spinor variables to the deformed position space correlator, we evaluate all undeformed spinors and the reference spinor $\tilde\eta$ on the Poincar\'e slice as $\tilde\Lambda_{i,\dot\alpha} =(1,w_i)$ and $\tilde\eta_{\dot\alpha} =(1,w_{\tilde\eta})$:  
\be
\braket{JJ\cdots J}(z)= (1+z)^{-2}\braket{J(w_1(z))J(w_2)\cdots J(w_n)}\,,
\label{equ:rescale}
\ee
where the overall factor accounts for the rescaling in defining the deformed spinor on the Poincar\'e slice.

\vskip 4pt
We expect that the correlator has singularities in $z$ only when two positions coincide, which is the OPE limit. The deformed correlator will therefore have singularities in $z$ if and only if $w_1(z_j)=w_j$, for any of the other fields $j=2,3,\cdots,n$. 
Using (\ref{equ:Def-w1}), this implies that the singularities are at position
\be
z_j=\frac{w_1-w_j}{w_j-w_{\tilde\eta}}\, .
\label{eq:defZi}
\ee
Equivalently, we get a singularity when the bracket $\langle 1(z_j)\hs j  \rangle $ vanishes, which reduces to $w_1(z)=w_j$ on the Poincar\'e slice due to equation \eqref{eqspinorbraketpoincareslice}. Using (\ref{equ:bracket}), this occurs at
\be
z_j=  \frac{\langle 1j\rangle}{ \langle j\tilde{\eta} \rangle}\, .
\ee
In particular, notice that the deformed correlator  in \eqref{eq-deformed-correlator} is not singular at $z=-1$ even though $w_1(z=-1)=\infty$, unless there is one vanishing deformed angle bracket $\braket{1(z=-1)j}=\braket{1j}-\braket{\tilde{\eta}j}=0$, or equivalently $z_j=-1$ (for some $j=2,3,\cdots,n$). Similarly, the deformed correlator is not singular at $z=\infty$ unless $z_j=\infty \iff \braket{j\tilde{\eta}}=0$ for some $j=2,3,\cdots,n$.\footnote{We can check this by writing the residue at $\infty$ as
	\be 
	\underset{z=\infty}{\rm Res}\left(\frac{\braket{JJ\cdots J}(z)}{z}\right)=-\underset{z'=0}{\rm Res}\left(z'\braket{J(\tilde\eta+z'\tilde\Lambda_1)J(\tilde\Lambda_2)\cdots J(\tilde\Lambda_n)}\right),\nonumber
	\ee
	which vanishes unless $\braket{j\tilde\eta}=0$ for some $j=2,3,\cdots,n$. } 
Thus, the poles that we are summing over in equation \eqref{eq-residue-sum} are always given by the OPE poles $z_I=z_j$ for $j=2,3,4,\cdots,n$. It may happen that one of these poles is $z=z_j=\infty$ for some $j$, in which case we need to compute its residue to obtain the correlator following \eqref{eq-residue-sum}. Remarkably, unlike what happens for the BCFW recursion relations applied to scattering amplitudes (recall Appendix \ref{sec:amplitudes-recursion}; see also, e.g.~\cite{Feng:2009ei} for a discussion of ``poles at infinity" in scattering amplitudes), we will be able to compute the residue at $z=z_j=\infty$ by using the OPE limit.

\subsubsection*{Residues from the OPE} We will use the OPE for conserved currents in the limit $z\to z_j$, or equivalently $w_1(z)\to w_j$, to find the residues for each singularity~$z=z_j$ (for $j=2,3,\cdots,n$). We use this input of dynamics to write residues of higher-point functions in terms of lower-point functions.
More specifically, writing explicitly the color indices $ a_j$ of the currents, we will use the following OPE of holomorphic currents:
\be
J_{ a_1}(w_1)J_{ a_j}(w_j) \sim \frac{k\hs\delta_{ a_1  a_j}}{(w_1-w_j)^2}+\frac{1}{(w_1-w_j)}i\hs f_{ a_1  a_j  c}\hs J_{ c}(w_j)\,, 
\label{equ:OPE-JJ}
\ee
where $k \in \mathbb{Z}_{>0}$ is the {\it level} and $f_{ a b c}$ are the {\it structure constants} of the current algebra. In \eqref{equ:OPE-JJ} and in subsequent equations, we use $\sim$ to signify that we only keep the divergent contributions to the OPE and discard regular terms.  
In terms of spinors, this OPE gets uplifted as
\be\label{eq-spinorope-JJ}
J_{ a_1}(\tilde\Lambda_1(z))J_{ a_j}(\tilde\Lambda_j) \sim \frac{k\hs\delta_{ a_1 a_j}}{\langle 1(z)j\rangle^2}+\frac{1}{\langle 1(z)j\rangle}\frac{\langle j\tilde\eta\rangle}{ \langle 1\tilde\eta\rangle}i\hs f_{ a_1 a_j  c}\hs J_{ c}(\tilde\Lambda_j)\,.
\ee
To prove that  \eqref{eq-spinorope-JJ} is equivalent to \eqref{equ:OPE-JJ}, we evaluate it on the Poincar\'e slice.
Using 
\be\label{equ:poincare-slice-sec3} \langle ij\rangle=\langle ij\rangle_P=w_{ij}\equiv w_i-w_j\,,
\ee 
we then get
\begin{align}
	J_{ a_1}(\tilde\Lambda_1(z))J_{ a_j}(\tilde\Lambda_j) &\sim  \frac{k\hs\delta_{ a_1 a_j}}{(w_{1j}+zw_{\tilde\eta j})^2}+\frac{1}{(w_{1j}+zw_{\tilde\eta j})}\frac{w_{j\tilde\eta }}{w_{1\tilde\eta}}i\hs f_{ a_1 a_j c}\hs J_{ c}(w_j) \nonumber \\
	&=  \frac{k\hs\delta_{ a_1 a_j}}{(w_{1j}+zw_{\tilde\eta j})^2}+\frac{(1+z)^{-1}}{(w_{1j}+zw_{\tilde\eta j})}i\hs f_{ a_1 a_j c}\hs J_{ c}(w_j) + {\rm regular}\nonumber \\
	&\sim \frac{1}{(1+z)^2}\left( \frac{k\hs\delta_{ a_1 a_j}}{(w_1(z)-w_j)^2}+\frac{i\hs f_{ a_1 a_j c}\hs J_{ c}(w_j)}{(w_1(z)-w_j)} \right) ,
\end{align}
where in the second line we replaced $w_{j\tilde\eta }/w_{1\tilde\eta} = (1+z_j)^{-1}$ (with $z_j$ as in \eqref{eq:defZi}) by $(1+z)^{-1}$, because the difference is regular at $z=z_j$. Taking into account the rescaling \eqref{equ:rescale}, we have therefore proven the equivalence of \eqref{eq-spinorope-JJ} and \eqref{equ:OPE-JJ} on the Poincar\'e slice. Since \eqref{eq-spinorope-JJ} is covariant under rescalings of the spinors $\tilde\Lambda_1$, $\tilde\Lambda_j$, and $\tilde\eta$, it holds on any slice.

\vskip 4pt

Using the OPE (\ref{eq-spinorope-JJ}), we can then compute the residues for the poles at  $z=z_j$ (for $j=2,3,4,\cdots,n$):
\begin{align}
	R_{z_j} &\equiv  \underset{z=z_j}{\rm Res} \left(\frac{\braket{J_{ a_1}(\tilde\Lambda_1(z))J_{ a_2}(\tilde\Lambda_2)\cdots J_{ a_n}(\tilde\Lambda_n)}}{z}\right) \\
	&=-\frac{k \hs\delta_{ a_1 a_j} }{\braket{1j}^2} \hs \braket{J_{ a_2}\cdots J_{ a_{j-1}} J_{ a_{j+1}} \cdots J_{ a_n}}-i\hs f_{ a_1   a_j   c}\hs \frac{\braket{j\tilde\eta}}{\braket{1j}\braket{1\tilde\eta}}\hs \braket{J_{  a_2}\cdots J_{  c}(\tilde\Lambda_j)\cdots J_{  a_n}}\,,\nonumber
\end{align}
where $J_{  a_j} \equiv J_{  a_j}(\tilde\Lambda_j)$.
It is easy to check that possible regular terms in the OPE limit, which were not written in \eqref{eq-spinorope-JJ}, do not contribute to this residue. Furthermore, notice that if we choose $\tilde\eta=\tilde\Lambda_j$ so that $\braket{j\tilde\eta}=0$, then the pole is at $z_j=\infty$, and there is no contribution proportional to the structure constant $f_{  a  b  c}$. As anticipated before, in contrast to the computation of residues of scattering amplitudes in the BCFW deformation, we are indeed able to compute the residue at $z=\infty$.

\subsubsection*{Recursion relation} Returning to \eqref{eq-residue-sum}, we sum all the residues we have just computed to find the undeformed $n$-point correlator in terms of lower-point correlators as a function of the embedding-space spinors:
\begin{align}\label{eqrecursionkacmoodygeneral}
		\braket{J_{ a_1}\cdots J_{ a_n}}
		&=-\sum_{j=2}^nR_{z_j}\\
		& =\sum_{j=2}^n \bigg(\frac{k \hs\delta_{ a_1 a_j} }{\braket{1j}^2} \hs \braket{J_{ a_2}\cdots J_{ a_{j-1}} J_{ a_{j+1}} \cdots J_{ a_n}} +i\hs f_{ a_1  a_j  c}\hs \frac{\braket{j\tilde\eta}}{\braket{1j}\braket{1\tilde\eta}}\hs \braket{J_{ a_2}\cdots J_{ c}(\tilde\Lambda_j)\cdots J_{ a_n}}\bigg).\nonumber
	\end{align}
	In this way, we can write any $n$-point function in terms of lower-point functions, with one and two fewer points. 
	The first term in (\ref{eqrecursionkacmoodygeneral}) corresponds to the disconnected part of the correlator when $n\geq 3$.
	Focusing on the connected part, we can write the recursion relation as
    \be 
	\braket{J_{ a_1}J_{ a_2}\cdots J_{ a_n}}_{\rm{c}}
	\ =\ \delta_{n2}\frac{k\hs \delta_{ a_1 a_2}}{\langle 12\rangle^2}+\sum_{j=2}^n \hs if_{ a_1  a_j  c}\hs \frac{\braket{j\tilde\eta}}{\braket{1j}\braket{1\tilde\eta}}\hs \braket{J_{ a_2}\cdots J_{ c}(\tilde\Lambda_j)\cdots J_{ a_n}}_{\rm{c}}\,,
	\label{equ:recursion-c}
	\ee
	which relates the connected $n$-point function to the connected $(n-1)$-point function. Note that the first term is nonzero only for $n=2$, while the second term is nonzero only for $n\geq3$.
	
	\vskip 4pt
	
	Note that these recursion relations, derived from the OPE between two conserved currents, are the same as those developed in the seminal papers \cite{Knizhnik:1984nr,Belavin:1984vu} when projected to the Poincar\'e slice. The main difference is that they have been uplifted to the space of spinors, where the symmetries are more manifest.
	
	\vskip 4pt

Using the recursion relation \eqref{equ:recursion-c} for the connected correlators, we can compute the two-point function, which just comes from the first term
	\be 
	\braket{J_{ a_1}J_{ a_2}} =\frac{k\hs \delta_{ a_1 a_2}}{\braket{12}^2}\,,
	\label{equ:YM2}
	\ee 
where we used that the one-point function $\braket{J_a}$ vanishes. Furthermore, using \eqref{equ:YM2}, we can compute the three-point function
		\be 
	\braket{J_{ a_1}J_{ a_2}J_{ a_3}} =\frac{-k\hs i\hs f_{ a_1 a_2 a_3}}{\braket{12}\braket{23}\braket{31}}\,,
\label{equ:YM3}		
		\ee 		
	 where the dependence on the reference spinor $\tilde\eta$ in the recursion cancels due to the Schouten identity\footnote{This identity is simply saying that three spinors $\tilde\Lambda_{i,\dot\alpha}$ in a two-dimensional vector space must be linearly dependent, and thus they must satisfy $\sum_{i=1}^3 b_i \tilde\Lambda_{i,\dot\alpha}=0$ for some non-trivial coefficients~$b_i$. Contracting this with $\epsilon^{\dot\alpha\dot \beta}\tilde\Lambda_{j,\dot\beta}$ for $j=1,2,3$, it is easy to deduce each $b_i$ up to an overall factor, which yields the Schouten identity
			\be 
\braket{12}\tilde\Lambda_{3,\dot\alpha}+\braket{23}\tilde\Lambda_{1,\dot\alpha}+\braket{31}\tilde\Lambda_{2,\dot\alpha}=0\,.
			\ee 
			This can be contracted with any other auxiliary spinor $\tilde\eta^{\dot\alpha}$ to get \eqref{eq-schouten}. 
			}
		\be \label{eq-schouten}
		\braket{12}\braket{3\tilde\eta}+\braket{23}\braket{1\tilde\eta}+\braket{31}\braket{2\tilde\eta}=0\,.
		\ee 
		We could therefore have chosen $\tilde\eta=\tilde\Lambda_2$ to simplify the calculation while obtaining the same result.  These two- and three-point functions coincide with our result in Section~\ref{sec:bootstrap} (for the case of conserved tensors with spin $|s|=1$), but the recursion fixes their 
		normalization to be consistent with the OPE in~(\ref{eq-spinorope-JJ}).\footnote{We may redefine the overall normalization of the currents as $\tilde J=g J$ with $g=1/\sqrt{k}$ to obtain a canonically normalized two-point function.\label{footnote-2pt-cft2}}

\subsection{Four-Point Function}

	Since the dynamics of the theory is specified by the OPE of holomorphic currents, we can use it to find a specific result for the four-point function. Indeed, using the recursion relation (\ref{equ:recursion-c}), we write the connected
	four-point function as
\begin{align}\label{eqfourpointgeneralkacmoody}
		\braket{J_{ a_1}J_{ a_2}J_{ a_3}J_{ a_4}}_{\rm{c}} &= i\hs f_{ a_1  a_2  c}\hs \frac{\braket{2\tilde\eta}}{\braket{12}\braket{1\tilde\eta}}\hs \braket{J_{ c}J_{ a_3} J_{ a_4}} +i\hs f_{ a_1  a_3  c}\hs \frac{\braket{3\tilde\eta}}{\braket{13}\braket{1\tilde\eta}}\hs \braket{J_{ a_2}J_{ c} J_{ a_4}} \nonumber\\
		&\quad +i\hs f_{ a_1  a_4  c}\hs \frac{\braket{4\tilde\eta}}{\braket{14}\braket{1\tilde\eta}}\hs \braket{J_{ a_2}J_{ a_3} J_{ c}}\,,
	\end{align}
where the various three-point functions are evaluated at positions $2,\,3$ and $4$.
Substituting the three-point function (\ref{equ:YM3}), we get
	\be
		\braket{JJJJ}_{\rm{c}} =\bigg(c_s\hs \frac{\braket{2\tilde\eta}}{\braket{12}\braket{1\tilde\eta}}+c_u\hs \frac{\braket{3\tilde\eta}}{\braket{13}\braket{1\tilde\eta}}+c_t\hs \frac{\braket{4\tilde\eta}}{\braket{14}\braket{1\tilde\eta}}\bigg) \frac{k}{\braket{23}\braket{34}\braket{42}}\,.
	\label{equ:YM4}
	\ee
	Here, the color indices in the correlator on the left-hand side are implicit, and we have defined the color factors
	\beq
	\begin{aligned}
		c_s&=f_{ a_1 a_2 c}f_{ a_3 a_4 c}\,,\\
		c_u&=f_{ a_3 a_1 c}f_{ a_2 a_4 c}\,,\\
        c_t&=f_{ a_1 a_4 c}f_{ a_2 a_3 c}\,, \label{eq:cscuct}
	\end{aligned}
	\eeq
	which satisfy the Jacobi identity $c_s+c_t+c_u=0$.
	It is easy to check that the result is invariant under permuting the fields $\{2,3,4\}$, and after a more involved computation it turns out to be invariant under any permutation of $\{1,2,3,4\}$, including those that involve the field $1$, thanks to the Jacobi identity. Thus, crossing symmetry is guaranteed by the Jacobi identity. 
	
	\vskip 4pt
	Defining the factors that multiply each color factor $c_s$, $c_t$, $c_u$ in \eqref{equ:YM4} as $N_s$, $N_t$, $N_u$, such that for example 
    \be 
    N_s =  \frac{\braket{2\tilde\eta}}{\braket{12}\braket{1\tilde\eta}}\frac{k}{\braket{23}\braket{34}\braket{42}}\,,
    \ee 
    we can write the four-point function as
	\be 
	\braket{JJJJ}_{\rm{c}}=c_s N_s+c_t N_t+ c_u N_u\,.
	\ee
	Due to the Jacobi identity, this expression is invariant under the so-called generalized gauge transformations
	\be\label{eq-generalizedgauge}
	\begin{aligned}
		N_s\mapsto&N_s+\alpha\,,\\
		N_t\mapsto&N_t+\alpha\,,\\
		N_u\mapsto&N_u+\alpha\,.
	\end{aligned}
	\ee
	In fact, changing the arbitrary spinor $\tilde\eta$ in the most generic way (while preserving~$\braket{1\tilde\eta}\neq 0$),
	\be 
	\tilde\eta\mapsto r_{\tilde\eta}(\tilde\eta+\beta \tilde\Lambda_1)\,,
	\ee 
	amounts to performing a generalized gauge transformation as in \eqref{eq-generalizedgauge}, with 
	\be 
	\alpha=\frac{-k\hs\beta}{\braket{23}\braket{34}\braket{42}\braket{1\tilde\eta}}\,.
	\ee
	In this way, the Jacobi identity ensures both crossing symmetry and that the four-point function does not depend on the choice of the arbitrary spinor $\tilde\eta$. 
	
	\vskip 4pt
	We can fix the gauge by choosing the reference spinor to be $\tilde\eta=\tilde\Lambda_2$, and thus the four-point function is written in this gauge as
	\be \label{equ:YM4-eta2}
	\braket{JJJJ}_{\rm{c}} = c_u \frac{(-k)}{\braket{12}\braket{34}\braket{13}\braket{42}}+c_t \frac{(-k)}{\braket{12}\braket{34}\braket{23}\braket{41}}\,,
	\ee
	which does not have a contribution in the $s$-channel. Although this correlator does not look symmetric under permutations, this is just a consequence of the gauge choice. Indeed, after projecting to the Poincar\'e slice using~\eqref{equ:poincare-slice-sec3} and imposing the Jacobi identity $c_s + c_t + c_u=0$ to reinstate $c_s$, we recover the connected part of the manifestly symmetric current four-point function that is displayed in equation~\eqref{eqJJJJap} in the appendix.

\subsection{Parke--Taylor Correlator}\label{ssec:parketaylor}
	
	In the same way that we computed the four-point function in terms of the three-point function, we can compute any  higher $n$-point function with $n>4$ using the recursion relation. Indeed, in this section, we will show using induction that the $n$-point function of conserved holomorphic currents can be written as	
	\be\label{eqnpointkacmoody}
\braket{J_{ a_1}J_{ a_2}\cdots J_{ a_n}}_{\rm{c}}=(-k)\displaystyle\sum_{\sigma\in S_{n-2}}(\tau_{ a_{\sigma_n}}\cdot \tau_{ a_{\sigma_{n-1}}} \cdots  \tau_{ a_{\sigma_3}})_{ a_1 a_2}~ G_n[12\sigma_3\cdots \sigma_n]\,,
\ee
	where the sum is over all permutations $\sigma:j\mapsto \sigma_j$ of the set $\{3,4,\cdots,n\}$, and the $(-k)$ factor is given by the normalization of the two-point function in \eqref{equ:YM2}. This factor is conventional; for instance, it becomes~$-g^{n-2}$ when we use the normalization of Footnote \ref{footnote-2pt-cft2}. Let us now define its main elements before delving into the proof.
	
	\vskip 4pt
	
	To begin with, the objects $\tau_b$ are defined as the matrices with elements $(\tau_{ b})_{ c a}=i\hs f_{ a b c}$ (i.e.~they are minus the generators of the Lie algebra in the adjoint representation). The product between these generators defines the possible color structures included in the previous formula
\begin{align} 
(\tau_{ a_{\sigma_n}}\cdot \tau_{ a_{\sigma_{n-1}}} \cdot \ldots \cdot  \tau_{ a_{\sigma_3}})_{ a_1 a_2}&=(\tau_{ a_{\sigma_n}})_{ a_1  c_n} (\tau_{ a_{\sigma_{n-1}}})_{ c_n  c_{n-1}} \cdot \ldots \cdot  (\tau_{ a_{\sigma_4}})_{ c_5 c_4}(\tau_{ a_{\sigma_3}})_{ c_4 a_2}\nonumber\\
&=i^{n-2}\hs f_{ c_{n} a_{\sigma_n} a_1}f_{ c_{n-1}  a_{\sigma_{n-1}} c_n}\cdots f_{ c_4  a_{\sigma_4} c_5} f_{ a_2  a_{\sigma_3} c_4}\,,\label{eq-ddm-basis}
\end{align}
where we obtained the second line simply using the definition of $(\tau_{b})_{{c}{a}}$ in terms of $f_{{a}{b}{c}}$.
These $(n-2)!$ color structures are the same as those used to decompose the gluon amplitude in YM theory in the DDM basis~\cite{DelDuca:1999iql,DelDuca:1999rs}.

	\vskip 4pt
	
	In addition, the so-called color-ordered correlator $G_n$ that is multiplying each color structure in \eqref{eqnpointkacmoody} is given simply by
	\be \label{eq-Parke--Taylor-correlator}
	G_n[12\sigma_3\cdots \sigma_n]=\frac{1}{\braket{12}\braket{2\sigma_3}\braket{\sigma_3\sigma_4}\cdots \braket{\sigma_n1}}\,.
	\ee 
For the trivial permutation $\sigma_j=j$, this is the analog of the Parke--Taylor amplitude \eqref{eq-Parke--Taylor-amplitudes},
 but for correlators of holomorphic conserved currents. Note that, while the Parke--Taylor amplitude corresponds to a helicity configuration with exactly two particles having negative helicity, the Parke--Taylor correlator in equation \eqref{eq-Parke--Taylor-correlator} corresponds to an all-plus configuration where all the currents are holomorphic $J_i=J^+_i$. This is why the latter formula \eqref{eq-Parke--Taylor-correlator} is given by just the denominator of the typical Parke--Taylor amplitude \eqref{eq-Parke--Taylor-amplitudes}. Connections and similarities between correlators of conserved currents in two-dimensional CFTs and flat-space scattering amplitudes of gluons in four dimensions have been observed in the literature, from the classic work~\cite{Nair:1988bq} of Nair to more recent works in celestial holography~\cite{He:2015zea,Pasterski:2017ylz}---see also~\cite{Bu:2023cef,Bu:2023vjt,Seet:2025mes}.
	
\vskip 4pt

Let us now turn to the proof of the Parke--Taylor correlator \eqref{eq-Parke--Taylor-correlator} by induction.
	
\paragraph{Proof} It is straightforward to check that the formula for the $n$-point correlator~\eqref{eqnpointkacmoody} with the color-ordered $G_n$ given by the Parke--Taylor correlator~\eqref{eq-Parke--Taylor-correlator} coincides exactly with the three-point function in \eqref{equ:YM3} for $n=3$. In order to complete the proof by induction, it only remains to show that this is true for $n$ provided it is true for $n-1$.

\vskip 4pt

The $n$-point correlator for $n\geq 3$ is given by the recursion relation in \eqref{equ:recursion-c}:
\be \label{eq-rec-eta2}
\braket{J_{ a_1}J_{ a_2}\cdots J_{ a_n}}_{\rm{c}}
\ =\ \sum_{j=3}^n (\tau_{ a_j})_{ a_1  c}\hs \frac{\braket{j2}}{\braket{j1}\braket{12}}\hs \braket{J_{ a_2}\cdots J_{ c}(\tilde\Lambda_j)\cdots J_{ a_n}}_{\rm{c}}\,,
\ee
where we have chosen the reference spinor to be $\tilde\eta=\tilde\Lambda_2$, and thus there is no contribution from the term with $j=2$. We can now use that the $(n-1)$-point correlator appearing in \eqref{eq-rec-eta2} is given by \eqref{eqnpointkacmoody}, namely
\be \label{eq-n-1-pt}
\braket{J_{ a_2}\cdots J_{ c}\cdots J_{ a_n}}_{\rm{c}}=(-k)\displaystyle\sum_{\sigma'\in S_{n-3}}(\tau_{ a_{\sigma_{n-1}'}}\cdots \tau_{ a_{\sigma_{3}'}})_{ c a_2}~ G_{n-1}[j2\sigma_3'\cdots \sigma_{n-1}']\,,
\ee 
where the sum is over the $(n-3)!$ permutations $\sigma'$ of the set $\{3,\cdots,j-1,j+1,\cdots,n\}$
without the element $j$, and the $(n-1)$-point color-ordered correlator is given by~\eqref{eq-Parke--Taylor-correlator}, i.e.
\be \label{eq-Gn-1}
G_{n-1}[j2\sigma_3'\cdots \sigma_{n-1}']=\frac{1}{\braket{j2}\braket{2\sigma_3'}\braket{\sigma_3'\sigma_4'}\cdots \braket{\sigma_{n-1}'j}}\,.
\ee 
Plugging the $(n-1)$-point correlator \eqref{eq-n-1-pt} in the right-hand side of \eqref{eq-rec-eta2}, we get
\be 
\braket{J_{ a_1}J_{ a_2}\cdots J_{ a_n}}_{\rm{c}}
\ =\ 
\sum_{j=3}^n \sum_{\sigma'\in S_{n-3}}(\tau_{ a_j}\cdot\tau_{ a_{\sigma_{n-1}'}}\cdots \tau_{ a_{\sigma_{3}'}} )_{ a_1  a_2}\hs \frac{\braket{j2}}{\braket{j1}\braket{12}}\hs G_{n-1}[j2\sigma_3'\cdots \sigma_{n-1}']\,.
\ee 
After reorganizing these two sums into only one sum over the $(n-2)!$ permutations $\sigma\in S_{n-2}$ of $\{3,4,\cdots,n\}$ defined as $\sigma_i=\sigma'_i$ for $3\leq i\leq n-1$ and $\sigma_n=j$, we can write \eqref{eq-rec-eta2} in the form of \eqref{eqnpointkacmoody} where the color-ordered correlator is given precisely by
\be\label{equ:recursion-Gn}
G_n[12\sigma_3\cdots \sigma_n]=\frac{\braket{\sigma_n2}}{\braket{\sigma_n1}\braket{12}}\hs G_{n-1}[\sigma_n 2\sigma_3\cdots \sigma_{n-1}]\,.
\ee
This yields a recursion relation for the color-ordered correlators $G_n$. Replacing $G_{n-1}$ by \eqref{eq-Gn-1}, it is trivial to obtain the Parke--Taylor correlator in \eqref{eq-Parke--Taylor-correlator}. This completes our proof.

\paragraph{Generalized gauge invariance}  Note that we solved the recursion \eqref{equ:recursion-c} by choosing the reference spinor to be $\tilde\eta=\tilde\Lambda_2$ in each step of the recursion. However, given the three-point function \eqref{equ:YM3}, we could have chosen different reference spinors $\tilde\eta_1,\cdots,\tilde\eta_{n-3}$ in each of the $n-3$ steps of the recursion. Let us now show by induction that any different choice will give the same result for the correlator, as expected.

\vskip 4pt

Notice that the recursion \eqref{equ:recursion-c} is invariant under $\tilde\eta\mapsto r_{\tilde\eta}(\tilde\eta+\beta \tilde\Lambda_{1})$ if and only if
\be \label{eq-eta-independent}
\sum_{j=2}^nf_{ a_1  a_j  c}\hs  \braket{J_{ a_2}\cdots J_{ c}(\tilde\Lambda_j)\cdots J_{ a_n}}_{\rm{c}}=0\,.
\ee 
This may be interpreted as the correlator being invariant under gauge transformations that would transform the color index of the current as $\delta J_{ a_j}=i\epsilon ^{a_1}\hs f_{ a_1  a_j  c}J_{ c}$ with $\epsilon^{a_1}$ the infinitesimal transformation parameter. In any case, we can prove that this is true by induction. To begin with, this statement is trivially correct when $n=3$ using the two-point function in \eqref{equ:YM2} and the anti-symmetry of $f_{ a  b c}$. Furthermore, it is a straightforward exercise to show that it is true for $n$ provided it holds for $n-1$. Indeed,
using this together with the recursion relations for the $(n-1)$-point correlator in \eqref{equ:recursion-c} and some algebraic manipulations, we can write the left-hand side of \eqref{eq-eta-independent} as 
\be 
\sum_{j=3}^n  \frac{(-1)\braket{j\tilde\eta}}{\braket{2j}\braket{2\tilde\eta}}(f_{ a_1  a_2  c} f_{ c  a_j  b}-f_{ a_2  a_j  c} f_{ a_1  c  b}+f_{ a_1  a_j  c}f_{ a_2  c  b})\braket{J_{ a_3}\cdots J_{ b}\cdots J_{ a_n}}_{\rm{c}}=0\,,
\ee 
which trivially vanishes due to the Jacobi identity. This proves that \eqref{eq-eta-independent} holds for all $n$, and thus that the recursion relations are independent of the choice of the reference spinor $\tilde\eta$ in each step. 

\vskip 4pt

As with the four-point function, changing the reference spinors $\tilde\eta_1,\cdots,\tilde\eta_{n-3}$ corresponds therefore to a generalized gauge transformation that leads to the same result for the correlator, albeit written as a different combination of color structures. Note that the color-ordered Parke--Taylor correlator~\eqref{eq-Parke--Taylor-correlator} is gauge invariant, because it was derived from the full correlator~\eqref{eqnpointkacmoody} expressed with $\tilde\eta=\tilde\Lambda_2$ in a minimal basis of $(n-2)!$ color structures that has no gauge redundancy left.

\subsection{Non-Abelian Chern--Simons in AdS\texorpdfstring{$_3$}{3}}

So far in this section, we have bootstrapped correlators of conserved spin-1 currents in 2d CFT using recursion relations similar to the BCFW recursion for scattering amplitudes. These relations are equivalent to the Ward identities obeyed by the correlators. We now show that, for spin-1 currents, the boundary correlators of non-abelian Chern--Simons theory in Euclidean AdS$_3$ satisfy the same Ward identities \cite{Witten:1988hf,Elitzur:1989nr, Moore:1989yh}. This identifies the bulk theory that reproduces the correlators obtained from the recursion relations.\footnote{We thank Jan de Boer for very useful discussions regarding Chern--Simons theory in EAdS$_3$.}

\subsubsection*{Action with a boundary term} Let us start with the bulk action of this theory:
\be
S_{\text{bulk}}=\frac{(-i)k}{4\pi}\int_M \text{Tr}\left(A\wedge dA+\frac{2}{3}\hs A\wedge A\wedge A\right),
\ee 
where $A=A^aT_a$ is the gauge field, which is a Lie-algebra-valued $1$-form,\footnote{We choose the conventions
 \be
 [T_a,T_b]=i\hs f_{abc}\hs T_c \,,\quad\quad \text{Tr}(T_aT_b)= \delta_{ab}\,,
 \ee 
 for the generators $T_a$ of the Lie algebra, with $a,b,c$ the color indices.} and the $(-i)$ factor is due to the Euclidean signature. We take the manifold $M$ to be Euclidean AdS$_3$, whose metric in Poincar\'e coordinates reads\footnote{The letter $z$ is used only in this subsection to denote the usual holographic coordinate of AdS, and should not be confused with the deformation parameter used elsewhere in this paper.}
\be 
ds^2=\frac{dz^2+dw\hs d\bar w}{z^2}\,.
\ee 
Here, the two-dimensional boundary is at $z\to 0$, and it is parameterized by the holomorphic and antiholomorphic coordinates $w=x^1+i x^2$, $\bar w=x^1-i x^2$. 

\vskip 4pt

Thus, we can show that the variation of the action is
\be 
\delta S_{\text{bulk}}=\frac{(-i)k}{2\pi}\int_M \text{Tr}\left(\delta A\wedge F\right) - \frac{(-i)k}{4\pi}\int_{\partial M}\text{Tr}\left(A \wedge \delta A\right),
\ee 
where we integrated by parts one term using Stokes' theorem, and defined the field strength $F=dA+A\wedge A$. After imposing the bulk equation of motion $F=0$, we obtain the following on-shell variation of the action \cite{deBoer:2013gz,deBoer:2014fra}:
\begin{align}
\delta S_{\text{bulk}}\big|_{\text{on-shell}}&=- \frac{(-i)k}{4\pi}\int_{\partial M}\text{Tr}\left(A \wedge \delta A\right)\nonumber\\
&=- \frac{(-i)k}{4\pi}\int dw\wedge d\bar w  ~ \text{Tr}\left(A_w \delta A_{\bar w}-A_{\bar w} \delta A_w\right) .
\end{align} 
Here, we used the coordinates $w,\bar w$ to parameterize the boundary, where the gauge field is $A=A_w \hs dw+A_{\bar w} \hs d\bar w$. Since Chern--Simons theory is first-order in derivatives, the equations of motion determine only one component of the boundary gauge field in terms of the other. Therefore, we can only impose a boundary condition on one of the two components, meaning we can set only one of the variations, $\delta A_{\bar w}$ or $\delta A_w$, to zero. Hence, the variational principle is not yet well defined as we cannot set boundary conditions such that the on-shell variation of the action vanishes. 

\vskip 4pt

To solve this issue, we must add a boundary term to the action that selects only one of the two variations. Let us consider boundary conditions where we fix $A_{\bar w}=a_{\bar w}$ on the boundary (i.e. we set $\delta A_{\bar w}=0$), and let us add the following boundary term to the action \cite{Elitzur:1989nr,Coussaert:1995zp,Kraus:2006wn}:
\be 
S_{\text{CS}}=S_{\text{bulk}}+S_{\text{bdy}}\,, \quad \text{where}\quad S_{\text{bdy}}=- \frac{(-i)k}{4\pi}\int dw\wedge d\bar w  ~ \text{Tr}\left(A_w \hs A_{\bar w}\right).
\ee 
The variation of the total action now reads
\be \label{equ:variation-offshell}
\delta S_{\text{CS}} =\frac{(-i)k}{2\pi}\int_M \text{Tr}\left(\delta A\wedge F\right)+\frac{k}{\pi}\int d^2x ~ \text{Tr}\left(A_w \delta A_{\bar w}\right) ,
\ee 
where we further used $dw\wedge d\bar w=-2i\hs d^2x$. This vanishes after going on-shell and imposing said boundary conditions, leading to a well-defined variational principle.

\subsubsection*{Gauge transformations} Furthermore, let us show how the action with this boundary term transforms under the following gauge transformations:
\be 
\delta_\beta A= d\beta+ [A,\beta]\,,
\ee 
where $\beta=\beta^a T_a$ belongs to the Lie algebra of the gauge group. We will see that this action is gauge invariant modulo a boundary term.

\vskip 4pt

Replacing the gauge variation
of the field in equation \eqref{equ:variation-offshell} leads to the following variation of the total action:\footnote{To show this explicitly, it is convenient to define the covariant exterior derivative $D$ applied to a Lie-algebra-valued $p$-form $\omega$ as
\be 
D\omega\equiv d\omega+A\wedge \omega-(-1)^p \omega\wedge A\,.
\ee
It is straightforward to prove that, for $\omega$ and $\eta$ a $p$-form and a $q$-form respectively, it satisfies
\be 
\text{Tr}\left(D\omega\wedge \eta\right)=d\left(\text{Tr}\left(\omega\wedge \eta\right)\right)-(-1)^p\hs\text{Tr}\left(\omega\wedge D\eta\right),
\ee 
and that $DF=0$ for the field strength $F=dA+A\wedge A$. This easily implies that
\be 
\text{Tr}\left(\delta_\beta A\wedge F\right)=\text{Tr}\left(D\beta\wedge F\right)=d\left(\text{Tr}\left(\beta\wedge F\right)\right).
\ee 
This total derivative yields the first term in \eqref{equ:deltabetaS-primera}.
}
\begin{equation}\label{equ:deltabetaS-primera}
\delta_\beta S_{\text{CS}} 
=\frac{(-i)k}{2\pi}\int_{\partial M} \text{Tr}\left(\beta\hs  F\right)- \frac{(-i)k}{2\pi}\int dw\wedge d\bar w  ~ \text{Tr}\left(A_w (\partial_{\bar w}\beta+[A_{\bar w},\beta])\right) .
\end{equation}
Further replacing
\be 
F\big|_{\partial M}=dw\wedge d\bar w\left(\partial_w A_{\bar w}-\partial_{\bar w} A_{w}+[A_w,A_{\bar w}]\right),
\ee 
using the cyclicity of the trace and integrating by parts yields
\be \label{equ:deltabetaS}
\delta_\beta S_{\text{CS}} =\frac{(-i)k}{2\pi}\int dw\wedge d\bar w~ \text{Tr}\left(\beta \hs \partial_{w}A_{\bar w}\right) .
\ee 
This variation trivially vanishes up to a boundary term, which will be important in the following derivation of the Ward--Takahashi identity.

\subsubsection*{Ward--Takahashi identity} Using the way the Chern--Simons action transforms under a gauge transformation, we will show that the corresponding boundary correlators of the dual currents satisfy a Ward--Takahashi identity.

\vskip 4pt

Consider the partition function of the Chern--Simons theory in Euclidean AdS$_3$, which is a functional of the chosen boundary condition $A_{\bar w}|_{\partial M}=a_{\bar w}$:
\be 
Z[a_{\bar w}]=\int_{A_{\bar w}|_{\partial M}=a_{\bar w}} \mathcal{D}A ~  e^{-S_{\text{CS}}[A]}\,.
\ee 
Its variation under the gauge transformation
\be 
\delta_\beta a_{\bar w}^a=\partial_{\bar w}\beta^a+ i\hs f^{abc} a_{\bar w}^b\beta^c\,,
\ee 
written in terms of the color components, is trivially
\be 
\delta_\beta \log Z[a_{\bar w}]=\frac{1}{Z}\hs \delta_\beta Z[a_{\bar w}]= \frac{1}{Z}\int_{A_{\bar w}|_{\partial M}=a_{\bar w}} \mathcal{D}A ~  e^{-S_{\text{CS}}[A]} \hs \left(-\delta_\beta S_{\text{CS}}\right) =-\langle \delta_\beta S_{\text{CS}} \rangle\,.
\ee 
Notice that the variation of the action does not vanish due to the boundary term in \eqref{equ:deltabetaS}. After using $dw\wedge d\bar w=-2 i\hs d^2x$, this implies
\be \label{equ:variation1}
\int d^2 x \hs\delta_\beta a_{\bar w}^a \hs\frac{\delta \log Z}{\delta a_{\bar w}^a }=\frac{k}{\pi} \int d^2 x ~ \beta^a \hs \partial_{w}a_{\bar w}^a \,.
\ee 
Since
\be 
\frac{\delta \log Z}{\delta a_{\bar w}^a }= \frac{1}{Z}\int_{A_{\bar w}|_{\partial M}=a_{\bar w}} \mathcal{D}A ~  e^{-S_{\text{CS}}[A]} \hs \left(-\frac{\delta S_{\text{CS}}}{\delta a_{\bar w}^a }\right)=-\frac{k}{\pi}\hs \langle A_w^a\rangle  \,,
\ee 
then equation \eqref{equ:variation1} amounts to
\be 
\int d^2 x \left(\beta^a\partial_{w} a_{\bar w}^a +\partial_{\bar w}\beta^a  \langle A_w^a\rangle +i\hs f^{abc} \langle A_w^a \rangle a_{\bar w}^b \beta^c\right)=0\,.
\ee 
After a straightforward integration by parts and relabeling of color indices, this implies
\be 
\partial_{w} a_{\bar w}^a -\partial_{\bar w}  \langle A_w^a\rangle+i\hs f^{abc}\langle A_w^b \rangle a_{\bar w}^c=0\,.
\ee 
Notice that this is essentially the equation of motion on the boundary $\langle F_{w \bar w}|_{\partial M} \rangle=0$.

\vskip 4pt

Defining the dual current in the boundary \cite{Witten:1998wy,Gukov:2004id}
\be 
J^a\equiv  k\hs  A_w^a \,,
\ee 
the functional derivative can be identified with the one-point function of the dual current in the presence of a source as
\be 
\langle J^a\rangle_{a_{\bar w}}\equiv - \pi \frac{\delta \log Z}{\delta a_{\bar w}^a} \,.
\ee 
Then we get
\be 
\partial_{\bar w}  \langle J^a\rangle_{a_{\bar w}}=k\hs \partial_{w} a_{\bar w}^a +i\hs f^{abc}\langle J^b \rangle_{a_{\bar w}} a_{\bar w}^c\,.
\ee 
Correlators of conserved currents are given by
\be 
\langle J^a J^{a_1}\cdots J^{a_n} \rangle=(-\pi )^{n+1}\frac{\delta^{n+1}\log Z}{\delta a_{\bar w}^a\delta a_{\bar w}^{a_1}\cdots \delta a_{\bar w}^{a_n}}\bigg|_{a_{\bar w}=0}=(-\pi )^{n}\frac{\delta^{n}\langle J^a\rangle_{a_{\bar w}}}{\delta a_{\bar w}^{a_1}\cdots \delta a_{\bar w}^{a_n}}\bigg|_{a_{\bar w}=0}.
\ee 
Then we take functional derivatives with respect to $a_{\bar w}^{a_i}$ evaluated at $a_{\bar w}=0$ and we get at two points
\be 
\partial_{\bar w}  \langle J^a J^{a_1}\rangle=(-\pi )\hs k\hs \partial_{w} \delta^{(2)}(w-w_1) \hs \delta^{aa_1}\,.
\ee 
At higher points, taking $(-\pi )^n$ times $n$ functional derivatives, we get
\begin{align}
\partial_{\bar w}  \langle J^aJ^{a_1}\cdots J^{a_n}\rangle&=\delta_{n1}(-\pi )\hs k\hs \partial_{w} \delta^{(2)}(w-w_1) \hs \delta^{aa_1} \nonumber\\
&\quad\quad+(-\pi )\sum_{j=1}^ni\hs f^{aba_j} \langle J^{a_1}\cdots J^{a_{j-1}} J^b J^{a_{j+1}}\cdots J^{a_n} \rangle \delta^{(2)}(w-w_j)\, ,
\end{align}
where in this equation each current $J^{a_i}$ is evaluated at location $w_i$.
This coincides precisely with the recursion relations obtained from the formalism of embedding-space spinors. Indeed, replacing $\langle ij\rangle=w_i-w_j$ in the recursion relation~\eqref{equ:recursion-c} for connected correlators, and taking a derivative with respect to $\bar w_1$ using the holomorphic anomaly (i.e.~$\partial_{\bar w}1/w=\pi\delta^{(2)}(w)$ and $\partial_{\bar w}1/w^2=-\pi\partial_w\delta^{(2)}(w)$)  yields exactly the same identity. Thus, the recursion obtained from a BCFW shift in said formalism is equivalent to the BPZ Ward--Takahashi identity satisfied by the Chern--Simons theory in the bulk AdS$_3$. For previous related work, see also \cite{Keranen:2014ava,Bhattacharya:2025udq}.

	\section{Correlators of Spin-2 Currents}

Following the procedure of the previous section, we can also obtain recursion relations for correlators of the holomorphic spin-2 stress-energy tensor at any number of points.\footnote{Spin-2 fields in the bulk are of course related to boundary gravitons. Like non-abelian gauge theory, their semiclassical dynamics is described by Chern--Simons theory \cite{Achucarro:1986uwr,Witten:1988hc}, with the boundary stress-energy tensor carrying a central charge controlled by Newton's constant \cite{Brown:1986nw}.} As for spin-1 currents, they will be an analog of the BCFW recursion relations, with the OPE between the corresponding fields being the input from dynamics that relates higher to lower points. Furthermore, we will find a way to double-copy the correlators of spin-$1$ currents to the correlators of spin-$2$ tensors at any number of points, which will yield a Parke--Taylor formula for correlators of the stress-energy tensor.

	\subsection{Recursion Relations}

	We consider the same complex deformation of the embedding-space spinor $\tilde\Lambda_1\mapsto \tilde\Lambda_1+z\tilde\eta$ as in the previous section, except that now we apply it to the $n$-point correlator of the holomorphic stress-energy tensor: 
	\be\label{eq-deformed-correlator-T}
	\braket{TT\cdots T}(z) =\braket{T(\tilde\Lambda_1(z))T(\tilde\Lambda_2)\cdots T(\tilde\Lambda_n)}\,.
	\ee
	As before, we assume that the deformed correlator is a meromorphic function of the deformation parameter $z$, and we write the undeformed correlator as a sum of residues:
	\begin{equation}\label{eq-residue-sum2}
		\braket{TT\cdots T}(z=0) =\oint \frac{dz}{2\pi i} \frac{\braket{TT\cdots T}(z)}{z}
		=-\sum_{i=2}^n \underset{z=z_i}{\rm Res}\left(\frac{\braket{TT\cdots T}(z)}{z}\right) ,
	\end{equation}
	where $z_i=\braket{1i}/\braket{i\tilde\eta}$, for $i=2,3,\cdots,n$.

	\paragraph{Residues}  To evaluate the residues in (\ref{eq-residue-sum2}), we use the OPE for the holomorphic stress-energy tensor, which in position space reads
	\be\label{equ:OPE-TT}
	T(w_1)T(w_i)\sim \frac{c/2}{(w_1-w_i)^4}+\frac{2 T(w_i)}{(w_1-w_i)^2}+\frac{1}{(w_1-w_i)}\frac{\partial}{\partial w_i} T(w_i)\,,
	\ee
	where $c$ is the central charge. In terms of spinors, this OPE can be written as
    	\begin{equation}\label{eq-spinorope-TT}
	T(\tilde\Lambda_1(z)) T(\tilde\Lambda_i) \sim  \frac{c/2}{\langle 1(z)i\rangle^4}+\left(\frac{1}{\langle 1(z)i\rangle} \frac{\langle i\tilde\eta\rangle}{\langle 1\tilde\eta\rangle}\right)^2 2 T(\tilde\Lambda_i) -\frac{1}{\langle 1(z)i\rangle} \frac{\langle i\tilde\eta\rangle^2}{\langle 1\tilde\eta\rangle^3}\tilde\eta_{\dot{\alpha}} \frac{\partial}{\partial \tilde\Lambda_{i,\dot{\alpha}}}T(\tilde\Lambda_i)\,.
	\end{equation}
	To prove that \eqref{eq-spinorope-TT} is equivalent to \eqref{equ:OPE-TT}, we project the former to the Poincar\'e slice. To evaluate the derivative operator on the Poincar\'e slice, we write $\tilde\Lambda_{i,\dot{\alpha}}=r_i(1,w_i)$ and use the chain rule 
	\be\nonumber\frac{\partial}{\partial\tilde\Lambda_{i,\dot{\alpha}}}=\frac{\partial r_i}{\partial \tilde\Lambda_{i,\dot{\alpha}}}\frac{\partial }{\partial r_i}+\frac{\partial w_i}{\partial \tilde\Lambda_{i,\dot{\alpha}}}\frac{\partial}{\partial w_i}\,.
	\ee 
	The derivative with respect to $r_i$ is then applied to $T(\tilde\Lambda_i)=r_i^{-4}\hs T(w_i)$, which scales as in \eqref{equ:LG}, with weight $-2h=-4$.
	Importantly, we only set $r_i=1$ \textit{after} taking the derivative with respect to $r_i$. Evaluating $\tilde\eta_{\dot{\alpha}}=(1,w_{\tilde\eta})$ on the Poincar\'e slice as well, we find 
	\be 
\tilde	\eta_{\dot{\alpha}} \frac{\partial}{\partial \tilde\Lambda_{i,\dot{\alpha}}}T(\tilde\Lambda_i) =-4 T(w_i)-w_{i\tilde\eta }\frac{\partial T(w_i)}{\partial w_i}\,.
	\ee
	Using $\langle ij\rangle =w_{ij}$ for the spinor brackets on the Poincar\'e slice, we can then project~\eqref{eq-spinorope-TT} as
	\begin{align}\label{eq-TT-middlestep}
T(\tilde\Lambda_1(z))T(\tilde\Lambda_i) &\sim  \frac{c/2}{(w_{1i}+z\hs w_{\tilde\eta i})^4}+ 2 \left(\frac{1}{(w_{1i}+zw_{\tilde\eta i})} \frac{w_{i\tilde\eta }}{w_{1\tilde\eta} }\right)^2 T(w_i)\nonumber\\
	&\quad\ -\frac{1}{(w_{1i}+zw_{\tilde\eta i})} \frac{w_{i\tilde\eta }^2}{w_{1\tilde\eta }^3}\left(-4 -w_{i\tilde\eta }\frac{\partial }{\partial w_i}\right)T(w_i) \,.
\end{align}
Here, we can substitute $w_{i\tilde\eta }/w_{1\tilde\eta} = 1/(1+z_i)$ and 
	\begin{align}
	\left(\displaystyle\frac{1+z}{1+z_i }\right)^2&\simeq 1+\frac{2(z-z_i)}{1+z_i}+\mathcal{O}((z-z_i)^2)\nonumber\\
	&\simeq  1+\displaystyle\frac{2}{w_{\tilde\eta i}} w_{1i}(z)+\mathcal{O}((z-z_i)^2)\,,
	\end{align}
	where we defined $w_{1i}(z)\equiv w_1(z)-w_i$. Further performing some algebraic manipulations while neglecting terms regular in $z-z_i$, we can write \eqref{eq-TT-middlestep} as
	\be 
	T(\tilde\Lambda_1(z))T(\tilde\Lambda_i)  \sim\frac{1}{(1+z)^4}\bigg(\frac{c/2}{w_{1i}(z)^4}+\frac{ 2T(w_i)}{w_{1i}(z)^2}+\frac{1}{w_{1i}(z)}\frac{\partial }{\partial w_i}T(w_i)  \bigg)\,.
	\ee 
	Taking into account the expected rescaling $(1+z)^{-4}$, we have therefore proven the equivalence of \eqref{eq-spinorope-TT} and \eqref{equ:OPE-TT} on the Poincar\'e slice. Since \eqref{eq-spinorope-TT} is covariant under rescalings of the spinors $\tilde\Lambda_1$, $\tilde\Lambda_i$, and $\tilde\eta$, it holds on any slice.

	\vskip 4pt
	Using the OPE in (\ref{eq-spinorope-TT}), we can compute the residues
	in \eqref{eq-residue-sum2}:
	\begin{align}
		R_{z_i} &\equiv  \underset{z=z_i}{\rm Res} \left(\frac{\braket{T(\tilde\Lambda_1(z))T_2\cdots T_n}}{z}\right) \\[4pt]
		&=-\frac{c/2}{\langle 1i\rangle^4}\braket{T_2\cdots T_{i-1}T_{i+1}\cdots T_n}-\Big[ 2 \left(\frac{1}{\langle 1i\rangle} \frac{\langle i\tilde\eta\rangle}{\langle 1\tilde\eta\rangle}\right)^2  -\frac{1}{\langle 1i\rangle} \frac{\langle i\tilde\eta\rangle^2}{\langle 1\tilde\eta\rangle^3}\tilde\eta_{\dot{\alpha}} \frac{\partial}{\partial \tilde\Lambda_{i,\dot{\alpha}}}\Big]\braket{T_2\cdots T_i\cdots T_n}\,,\nonumber
	\end{align}
where we introduced the shorthand $T_{j} \equiv T(\tilde\Lambda_j)$.
	It is easy to check that possible regular terms in the OPE limit, which were not written in \eqref{eq-spinorope-TT}, do not contribute to this residue. Notice that if we choose $\tilde\eta=\tilde\Lambda_i$, so that $\braket{i\tilde\eta}=0$, then we get a pole at infinity, $z_i=\infty$, and the only contribution is that proportional to the central charge $c$.

	\paragraph{Recursion relation} Returning to \eqref{eq-residue-sum2}, we sum all the residues to find the undeformed $n$-point correlator in terms of lower-point correlators as a function of the embedding-space spinors:
	\begin{align}\label{eqrecursionT}
		\braket{T_{1}\cdots T_{n}}
		&=-\sum_{i=2}^nR_{z_i}\nonumber\\
		&=\sum_{i=2}^n \bigg(\frac{c/2}{\langle 1i\rangle^4}\braket{T_2\cdots T_{i-1}T_{i+1}\cdots T_n}+\left(\frac{1}{\langle 1i\rangle} \frac{\langle i\tilde\eta\rangle}{\langle 1\tilde\eta\rangle}\right)^2 2 \braket{T_2\cdots  T_n}\nonumber\\
		&\quad\quad -\frac{1}{\langle 1i\rangle} \frac{\langle i\tilde\eta\rangle^2}{\langle 1\tilde\eta\rangle^3}\tilde\eta_{\dot{\alpha}} \frac{\partial}{\partial\tilde\Lambda_{i,\dot{\alpha}}}\braket{T_2 \cdots T_n}\bigg)\,.
	\end{align}
	In this way, we can write any $n$-point function in terms of lower-point functions, with one and two fewer points. 
	The first term in (\ref{eqrecursionT}) corresponds to the disconnected part of the correlator when $n\geq 3$.
	Focusing on the connected part, we can write the recursion relation as
	\be 
	\braket{T_{1}\cdots T_{n}}_{\rm{c}}
=\delta_{n2}\frac{c/2}{\langle 12\rangle^4}+\sum_{i=2}^n \left(2\left(\frac{1}{\langle 1i\rangle} \frac{\langle i\tilde\eta\rangle}{\langle 1\tilde\eta\rangle}\right)^2-\frac{1}{\langle 1i\rangle} \frac{\langle i\tilde\eta\rangle^2}{\langle 1\tilde\eta\rangle^3}\tilde\eta_{\dot{\alpha}} \frac{\partial}{\partial \tilde\Lambda_{i,\dot{\alpha}}} \right)\braket{T_{2}\cdots T_{n}}_{\rm{c}}\,,
	\label{equ:recursion-c2}
	\ee
	which relates the connected $n$-point function to the connected $(n-1)$-point function.

\vskip 4pt

Let us prove that this recursion does not depend on the choice of the reference spinor $\tilde\eta$. 
	Note that the variation of the right-hand side of \eqref{equ:recursion-c2} under the shift~$\tilde\eta\mapsto r_{\tilde\eta}(\tilde\eta+\beta \tilde\Lambda_1)$, up to first order in $\beta$, is given by the left-hand side of 
\be 
\frac{3}{\braket{1\tilde\eta}^3}\tilde\eta_{\dot\beta}\tilde\eta_{\dot\alpha}\sum_{i=2}^nJ_i^{\dot\alpha\dot\beta}\braket{T_{2}\cdots T_{n}}_{\rm{c}}=0\,,
\ee 
where we used that the $(n-1)$-point correlator is a homogeneous function of the spinor $\tilde\Lambda_{i,\dot\alpha}$ of degree $-2h=-4$ together with some algebraic manipulations. This expression vanishes because the $(n-1)$-point correlator is invariant under the (global) conformal transformations generated by the operators
\be 
J_i^{\dot\alpha\dot\beta}=\tilde\Lambda_i^{(\dot{\alpha}}\frac{\partial}{\partial \tilde\Lambda_{i,\dot{\beta})}}\,,
\ee 
introduced in equation \eqref{eq-cft2-generators}. Hence, the recursion relations \eqref{equ:recursion-c2} are invariant under infinitesimal modifications of the reference spinor $\tilde\eta$ since we proved this to first order in $\beta$. Of course, exponentiating this infinitesimal transformation, we would get finite transformations $\tilde\eta\mapsto \tilde\eta+\beta \tilde\Lambda_1$ with $\beta$ not infinitesimal, and these must leave the recursion relation invariant as well.

\vskip 4pt

As before, these recursion relations, derived from the $TT$ OPE, are the same as those developed in \cite{Knizhnik:1984nr,Belavin:1984vu} when projected to the Poincar\'e slice. The main novelty is that they have been uplifted to the space of spinors, where the symmetries are more manifest.

\subsection{Four-Point Function}

We will compute the two-, three- and four-point functions of the stress-energy tensor using these recursion relations. However, before delving into this calculation, it will be useful for practical applications to write the recursion relation (\ref{equ:recursion-c2}) in a slightly different form. Indeed, using
	\be 
	\tilde\eta_{\dot{\alpha}}\frac{\partial}{\partial \tilde\Lambda_{i,\dot{\alpha}}}\braket{T_2\cdots T_n}=\sum_{k=2,\hs  k\neq i}^n \braket{k\tilde\eta}\frac{\partial}{\partial (\braket{ki})}\braket{T_2\cdots T_n}\,,
	\ee
	together with the usual Schouten identity \eqref{eq-schouten} for these spinors and some algebraic manipulations, allows us to rewrite the recursion relation for $n\geq 3$ as
	\begin{align} 
		\braket{T_{1}\cdots T_{n}}_{\rm{c}}
		& = \sum_{i=3}^n\sum_{k=2}^{i-1}\bigg[\frac{2}{(n-2)} \left(\frac{\langle ki \rangle}{\langle 1i\rangle\langle 1k\rangle}\right)^2+\nonumber\\
		&\quad \quad + \frac{1}{\braket{1\tilde\eta}^2} \frac{\langle i\tilde\eta\rangle}{\langle 1i\rangle} \frac{\langle k\tilde\eta\rangle}{\langle 1k\rangle}\left(\frac{4}{(n-2)}+\braket{ki}\frac{\partial}{\partial (\braket{ki})} \right)\bigg]\braket{T_{2}\cdots T_{n}}_{\rm{c}}\,.
		\label{equ:recursion-c3}
	\end{align}
	This form of the recursion relation is easier to use in practice because the correlators are just a sum of terms that are homogeneous in each spinor bracket $\braket{ki}$, and it is then simple to perform the partial derivatives. 
	
	\vskip 4pt
	
Starting from the recursion relation \eqref{equ:recursion-c2} as well as using the fact that the one-point function~$\braket{T}$ vanishes, we can obtain the two-point function  	
\be 
	\braket{T_1T_2} =\frac{c/2}{\braket{12}^4}\,.
	\label{equ:T2}
	\ee 
	Next, using this two-point function as a seed for the form \eqref{equ:recursion-c3} of the recursion, we can obtain the three-point function
	\be 
		\braket{T_1T_2T_3} =\frac{c}{\braket{12}^2\braket{23}^2\braket{31}^2}\,.
	\label{equ:T3}
	\ee 
	These correlators coincide with our result in Section~\ref{sec:bootstrap} (for the case of conserved tensors with spin $|s|=2$) and fix the normalization of both of them to be consistent with the OPE in~(\ref{eq-spinorope-TT}).

\vskip 4pt

Finally, we can compute the connected four-point function from the three-point function \eqref{equ:T3} by using the recursion relation \eqref{equ:recursion-c3} with $n=4$. It is straightforward to check that, due to the scaling of the three-point function \eqref{equ:T3} with respect to each spinor bracket, the second line in the recursion \eqref{equ:recursion-c3}  will vanish for the case $n=4$. Hence, we get the following simple expression for the connected four-point function:
	\begin{align}
		\braket{T_1T_2T_3T_4}_{\rm{c}}&=\left(\sum_{i=3}^4\sum_{k=2}^{i-1}\left(\frac{\langle ki \rangle}{\langle 1i\rangle\langle 1k\rangle}\right)^2\right)\braket{T_2T_3T_4}\nonumber\\
		&=\left(\left(\frac{\langle 23 \rangle}{\langle 13\rangle\langle 12\rangle}\right)^2+\left(\frac{\langle 34 \rangle}{\langle 14\rangle\langle 13\rangle}\right)^2+\left(\frac{\langle 24 \rangle}{\langle 14\rangle\langle 12\rangle}\right)^2\right)\frac{c}{\braket{23}^2\braket{34}^2\braket{42}^2}\nonumber\\
		&=\frac{c}{\braket{12}^2\braket{31}^2\braket{34}^2\braket{42}^2}+\frac{c}{\braket{31}^2\braket{41}^2\braket{23}^2\braket{42}^2}+\frac{c}{\braket{12}^2\braket{41}^2\braket{34}^2\braket{23}^2}\,.
	\end{align}
	After projecting to the Poincar\'e slice, this result coincides with the known connected four-point function that we can obtain from the position-space Ward identity 
\be 
	\braket{T(w_1) T(w_2) T(w_3) T(w_4)}_{\rm{c}} = \sum_{j=2}^4  \left[\frac{2}{w_{1j}^2} + \frac{1}{w_{1j}} \frac{\partial}{\partial w_j} \right]\hspace{-0.025cm} \braket{T(w_2) T(w_3) T(w_4)}  \,,
\ee 
where we again used the definition $w_{1j}\equiv w_1-w_j$.

\subsection{Double Copy}

It is easy to see that the recursion relation \eqref{equ:recursion-c2} for the stress-energy tensor correlators can be equivalently written in such a way that each summand has only one term as
\be 
\braket{T_{1}\cdots T_{n}}_{\rm{c}}
=\sum_{i=2}^n \hat\tau_i^{\hs 1\tilde\eta}
\left(\left( \frac{\langle i\tilde\eta\rangle}{\langle i1\rangle\langle 1\tilde\eta\rangle}\right)^2\braket{T_{2}\cdots T_{n}}_{\rm{c}}\right),
\label{equ:recursion-c4}
\ee
where we defined the differential operator
\be 
\hat\tau_i^{\hs j\tilde\eta}\equiv \frac{\braket{ij}}{\braket{j\tilde\eta}}\tilde\eta_{\dot{\alpha}} \frac{\partial}{\partial \tilde\Lambda_{i,\dot{\alpha}}}=\frac{\tilde\Lambda_{j,\dot\beta} \tilde\eta_{\dot\alpha}}{\langle j\tilde\eta\rangle }\hs F_i^{\dot\beta\dot\alpha}
\,,
\ee 
with $F_i^{\dot\beta\dot\alpha}$ defined in equation~\eqref{equ:def-operator-F}, which does not scale with respect to any spinor. Note that this form of the recursion is reminiscent of the one of conserved currents in \eqref{eq-rec-eta2}. Indeed, it is tantalizing to think of a ``double-copy" (in the spirit of \cite{Bern:2008qj, Bern:2010ue}) procedure that involves mapping the structure constants to this differential operator as
\be
(\tau_{ a_i})_{ a_1  c}\mapsto \hat\tau_i^{\hs1\tilde\eta}\,,
\ee
and ``squaring" the kinematic numerators in some way. We will see that this is indeed possible at any number of points.\footnote{Color-kinematics duality and the double copy have been explored in a variety of contexts in curved spacetime. For a small sample of works in this direction see \cite{Lipstein:2019mpu,Armstrong:2020woi, Albayrak:2020fyp,Alday:2021odx,Diwakar:2021juk,Zhou:2021gnu,Herderschee:2022ntr,CarrilloGonzalez:2022ggn,Cheung:2022pdk,Lee:2022fgr,Beetar:2024ptv,CarrilloGonzalez:2026phk}.}

\paragraph{Two and three points} It is straightforward to double-copy the two-point function by squaring the kinematic numerator of $\braket{J_{ a_1}J_{ a_2}}$ in \eqref{equ:YM2} (i.e.~dropping only the color factor $\delta_{ a_1  a_2}$ and keeping the constant $k$), which yields the correct two-point function $\braket{T_1T_2}$ in \eqref{equ:T2} provided we replace $k^2$ with $c/2$. 

\vskip 4pt

If we naively apply this procedure to the three-point function, the result of dropping the structure constants and squaring the three-point function of conserved currents
\be \label{equ:YM3-again}
\braket{J_{ a_1}J_{ a_2}J_{ a_3}} =\frac{-k\hs i\hs f_{ a_1 a_2 a_3}}{\braket{12}\braket{23}\braket{31}}\,,
\ee 
would coincide with the three-point function of the stress-energy tensor in \eqref{equ:T3} provided we replace $k^2$ by $c$. However, this replacement differs from the one for the two-point function by a factor of $2$. We will therefore perform a different double-copy procedure that involves replacing the color factor $(\tau_{a_2})_{a_3a_1}=i\hs f_{a_1a_2a_3}$ by the appropriate differential operators $\hat \tau$ instead of dropping them, which will work with the correct replacement $k^2\mapsto c/2$. Remarkably, the latter procedure generalizes to higher points.

\vskip 4pt

 Using the form \eqref{equ:recursion-c4} of the recursion, with $\tilde\eta=\tilde\Lambda_2$, we can write the three-point correlator as
\begin{align}\label{equ:T3-tauhat}
\braket{T_1T_2T_3}&=\hat\tau_3^{\hs12}\left(\left(\frac{\braket{32}}{\braket{31}\braket{12}}\right)^2\langle T_2T_3\rangle\right)\nonumber\\
&=\frac{c}{2}~\hat\tau_3^{\hs12}\left(\frac{1}{\braket{31}\braket{12}\braket{23}}\right)^2,
\end{align}
where we used the two-point function \eqref{equ:T2} as an input. Thus, we may say that the double copy of the three-point function of conserved currents in \eqref{equ:YM3-again} consists in replacing the structure constant by this differential operator as 
\be 
i\hs f_{ a_1  a_2  a_3}=(\tau_{ a_3})_{ a_1  a_2}\mapsto \hat\tau_3^{\hs12}\,,
\ee 
applying it to the square of the kinematic numerator of \eqref{equ:YM3-again} (where we square both the angle brackets and the constant $-k$), and performing the replacement $k^2\mapsto c/2$ consistent with the double copy of the two-point function.

\vskip 4pt

It turns out that applying this differential operator to the adjacent object in \eqref{equ:T3-tauhat} simply yields a factor of $2$, leading to the expected result for the three-point correlator \eqref{equ:T3} of the stress-energy tensor
\be 
\braket{T_1T_2T_3}=\frac{c}{\left(\braket{31}\braket{12}\braket{23}\right)^2}\,.
\ee 
Moreover, the result is the same if we permute the three particles in the operator $\hat\tau_3^{\hs12}$ (i.e.~if we replace it with $\hat\tau_{\sigma_3}^{\hs\sigma_1\sigma_2}$, where $\sigma\in S_3$ is any permutation of $\{1,2,3\}$).

\paragraph{Four points} Let us now see how to extend this double-copy procedure to four points. Using the recursion \eqref{equ:recursion-c4}, we can write the four-point correlator as
\be \label{equ:recursion-TTTT}
\braket{T_1T_2T_3T_4}_{\rm{c}}=\sum_{i=2}^4\hat\tau_i^{\hs1\tilde\eta}\left(\left( \frac{\langle i\tilde\eta\rangle}{\langle i1\rangle\langle 1\tilde\eta\rangle}\right)^2\braket{T_{2}T_3 T_{4}}\right).
\ee 
Further expressing the three-point function as in \eqref{equ:T3-tauhat},
\be \label{equ:TTT-tauijk}
\braket{T_{2}T_3 T_{4}}=\frac{c}{2}~\hat\tau_l^{\hs jk}\left(\frac{1}{\braket{23}\braket{34}\braket{42}}\right)^2,
\ee 
where $\{l,j,k\}$ is a permutation of $\{2,3,4\}$, we can write the four-point correlator as the sum of three terms:
\be \label{equ:TTTT-doublecopy}
\braket{T_1T_2T_3T_4}_{\rm{c}}=\frac{c}{2}\hs\hat\tau_2^{\hs1\tilde\eta}\cdot\hat{\tau}_{3}^{\hs24}\cdot \left(\frac{\langle 2\tilde\eta\rangle}{\langle 21\rangle\langle 1\tilde\eta\rangle}\frac{1}{\braket{23}\braket{34}\braket{42}}\right)^2+\text{cyclic}(2,3,4)\,.
\ee 
Here, we exchanged $\hat\tau_l^{\hs jk}$ with the angle brackets in \eqref{equ:recursion-TTTT} assuming that $\tilde\eta$ is independent of each $\tilde\Lambda_l$, with $l=2,3,4$ in each term, and summed over cyclic permutations of $\{2,3,4\}$. If this assumption is not true, we may always replace $\hat\tau_l^{\hs jk}$ by $\hat{\tau}_{k}^{\hs jl}$ in \eqref{equ:TTT-tauijk}, and we are then allowed to exchange $\hat\tau_{k}^{\hs jl}$ with the angle brackets in \eqref{equ:recursion-TTTT}.

\vskip 4pt

Notice that applying this double-copy procedure to the four-point function \eqref{equ:YM4} of conserved currents 
	\be
	\braket{JJJJ}_{\rm{c}} =(-c_s)\hs\left( \frac{\braket{2\tilde\eta}}{\braket{21}\braket{1\tilde\eta}}\frac{k}{\braket{23}\braket{34}\braket{42}}\right) +\text{cyclic}(2,3,4)\,,
\ee
yields precisely the four-point function \eqref{equ:TTTT-doublecopy} of the stress-energy tensor that we just computed. Indeed, this simply amounts to replacing the color structures as
\be 
-c_s=(\tau_{ a_2}\cdot \tau_{ a_3})_{ a_1  a_4}\mapsto \hat\tau_2^{\hs 1\tilde\eta}\cdot\hat{\tau}_{3}^{\hs 24}\,,
\ee 
and similarly for the other channels, squaring the kinematic numerators, and replacing at the end $k^2$ by $c/2$. This yields trivially the correct four-point function~\eqref{equ:TTTT-doublecopy}.

\vskip 4pt

Note that the reference spinor $\tilde\eta$ is arbitrary here. We may therefore perform the same double-copy procedure, but with a specific choice, like $\tilde\eta=\tilde\Lambda_2$ for instance. It is straightforward to show that the recursion \eqref{equ:recursion-TTTT}, for this choice of $\tilde\eta$, yields the following four-point function
\begin{align} \label{equ:TTTT-eta2}
\braket{T_1T_2T_3T_4}_{\rm{c}}&=\frac{c}{2}\hs\hat\tau_3^{\hs 12}\cdot \hat\tau_4^{\hs 32}\cdot \left(\frac{1}{\braket{12}\braket{24}\braket{43}\braket{31}}\right)^2+\frac{c}{2}\hs\hat\tau_4^{\hs 12}\cdot \hat\tau_3^{\hs 42}\cdot \left(\frac{1}{\braket{12}\braket{23}\braket{34}\braket{41}}\right)^2.
\end{align}
This can be obtained by double-copying the four-point function of conserved currents written in the choice $\tilde\eta=\tilde\Lambda_2$, which was computed in \eqref{equ:YM4-eta2}:
\be 
	\braket{JJJJ}_{\rm{c}} =(-c_u)\hs\left( \frac{k}{\braket{12}\braket{24}\braket{34}\braket{31}}\right) +c_t\hs\left( \frac{(-k)}{\braket{12}\braket{23}\braket{34}\braket{41}}\right).
\ee 
To be explicit, this procedure consists in replacing the color structures with the appropriate differential operators as
\begin{align}
-c_u&=(\tau_{ a_3}\cdot \tau_{ a_4})_{ a_1  a_2}\mapsto \hat\tau_3^{\hs 12}\cdot \hat\tau_4^{\hs 32}\,,\\
c_t&=(\tau_{ a_4}\cdot \tau_{ a_3})_{ a_1  a_2}\mapsto \hat\tau_4^{\hs 12}\cdot \hat\tau_3^{\hs 42}\,,
\end{align}
squaring the kinematic numerators, and replacing $k^2$ by $c/2$. Indeed, this trivially yields the correct four-point function in \eqref{equ:TTTT-eta2}.

\paragraph{Arbitrary points}Finally, let us extend this to any number of points. Applying this double-copy procedure to the $n$-point function of holomorphic conserved currents in \eqref{eqnpointkacmoody}:
\be\label{eqnpointkacmoody-again}
\braket{J_{ a_1}J_{ a_2}\cdots J_{ a_n}}_{\rm{c}}=(-k)\displaystyle\sum_{\sigma\in S_{n-2}}(\tau_{ a_{\sigma_n}}\cdot  \cdots  \tau_{ a_{\sigma_3}})_{ a_1 a_2}~ G_n[12\sigma_3\cdots \sigma_n]\,,
\ee
written in the gauge $\tilde\eta=\tilde\Lambda_2$, amounts to replacing the color structures by the differential operators
\be 
(\tau_{ a_{\sigma_n}}\cdot \tau_{ a_{\sigma_{n-1}}} \cdots  \tau_{ a_{\sigma_3}})_{ a_1 a_2}\mapsto \hat\tau_{\sigma_n}^{\hs 12}\cdot \hat\tau_{\sigma_{n-1}}^{\hs \sigma_n2}\cdots \hat\tau_{\sigma_3}^{\hs \sigma_42}
\,,
\ee 
squaring the kinematic numerators, and replacing $k^2$ by $c/2$. In particular, the last two steps are equivalent to directly replacing 
\be 
(-k)G_n\mapsto \frac{c}{2}\hs G_n^2\,,
\ee
where we recall that the color-ordered correlator is given by \eqref{eq-Parke--Taylor-correlator}:
\be \label{eq-Parke--Taylor-correlator-again}
G_n[12\sigma_3\cdots \sigma_n]=\frac{1}{\braket{12}\braket{2\sigma_3}\braket{\sigma_3\sigma_4}\cdots \braket{\sigma_n1}}\,.
\ee 
Putting everything together, the result of this double-copy procedure is the following correlator
\be\label{eqnpointT}
\braket{T_{1}T_{2}\cdots T_{n}}_{\rm{c}}=\frac{c}{2}\displaystyle\sum_{\sigma\in S_{n-2}}(\hat\tau_{\sigma_n}^{\hs 12}\cdot \hat\tau_{\sigma_{n-1}}^{\hs \sigma_n2}\cdots \hat\tau_{\sigma_3}^{\hs \sigma_42})~ (G_n[12\sigma_3\cdots \sigma_n])^2\,,
\ee
where the sum is over all $(n-2)!$ permutations $\sigma:j\mapsto \sigma_j$ of $\{3,4,\cdots,n\}$. This yields precisely the $n$-point function of holomorphic stress-energy tensors obtained from the recursion \eqref{equ:recursion-c4}.

\vskip 4pt

It is straightforward to prove this by induction following the same steps as in Section~\ref{ssec:parketaylor}. Indeed, we already know that this formula yields the correct three-point function in \eqref{equ:T3-tauhat} for $n=3$. To prove that it holds for $n$ given that it holds for $n-1$, we can easily use the recursion \eqref{equ:recursion-c4} in the gauge $\tilde\eta=\tilde\Lambda_2$, and then use the formula \eqref{eqnpointT} for the $(n-1)$-point correlator. Performing the same steps as in Section \ref{ssec:parketaylor}, this yields
\begin{equation}
\braket{T_{1}\cdots T_{n}}_{\rm{c}}=\frac{c}{2}\sum_{\sigma\in S_{n-2}} \hat\tau_{\sigma_n}^{\hs12}\bigg[\left( \frac{\langle \sigma_n 2 \rangle}{\langle \sigma_n1\rangle\langle 12 \rangle}\right)^2 ( \hat\tau_{\sigma_{n-1}}^{\hs \sigma_n2}\cdots \hat\tau_{\sigma_3}^{\hs \sigma_42})(G_{n-1}[\sigma_n2\sigma_3\cdots \sigma_{n-1}])^2\bigg].
\end{equation}
Since the angle brackets trivially commute with the differential operators $( \hat\tau_{\sigma_{n-1}}^{\hs \sigma_n2}\cdots \hat\tau_{\sigma_3}^{\hs \sigma_42})$, we can express this $n$-point correlator in the form of equation \eqref{eqnpointT}, where the corresponding $(G_n)^2$ satisfies the recursion
\be 
(G_n[12\sigma_3\cdots \sigma_n])^2=\left( \frac{\langle \sigma_n 2 \rangle}{\langle \sigma_n1\rangle\langle 12 \rangle}\right)^2 (G_{n-1}[\sigma_n2\sigma_3\cdots \sigma_{n-1}])^2\,.
\ee 
This is precisely the square of the recursion \eqref{equ:recursion-Gn} for the color-ordered correlators~$G_n$. We thus conclude that $G_n^2$ is given by the square of \eqref{eq-Parke--Taylor-correlator-again}. This therefore shows that the $n$-point correlator of stress-energy tensors is given by~\eqref{eqnpointT}, obtained as the double copy of the $n$-point correlator of conserved currents.

\vskip 4pt

The fact that we found a double-copy structure for correlators at any number of points may not seem surprising given the Sugawara construction \cite{Sugawara:1968gq}. This consists in building a stress-energy tensor from the normal-ordered product of two holomorphic conserved currents as
\be \label{eq-sugawara}
T(w)=\gamma :J_{ a}J_{ a}:(w)\equiv \gamma \lim_{w'\to w}\Big[J_{ a}(w')J_{ a}(w)-\langle J_{ a}(w')J_{ a}(w)\rangle \Big]\,,
\ee 
where we simply subtracted all the singular terms in the OPE, which is the definition of the normal-ordered product~\cite{DiFrancesco:1997nk}.\footnote{Note that the singular terms in the $J_{a}J_{ a}$ OPE \eqref{equ:OPE-JJ} in this case coincide precisely with the two-point function $\langle J_{ a}(w')J_{ a}(w)\rangle= ~(k\hs \dim(g) )/(w'-w)^2$, where $\dim(g)=\delta_{aa}$ is the dimension of the Lie algebra. This is because we consider the same color index for both currents, and thus the other singular term with the structure constant is proportional to $f_{aab}$ for some $b$ and vanishes due to the antisymmetric nature of structure constants.} Here, $\gamma=(2(k+h^{\vee}))^{-1}$, where $h^{\vee}$ is the dual Coxeter number, defined by $f_{abc}f\indices{_{d}^{bc}}=2h^{\vee}\hs \delta_{ad}$. This value for the proportionality constant $\gamma$ ensures precisely the expected $TT$ OPE~\eqref{equ:OPE-TT}, and thus the latter can be derived from the $JJ$ OPE via the Sugawara construction. Hence, ultimately it should be possible to derive the stress-energy tensor correlators from the current correlators with this construction, because their recursion relations are based on their OPEs.  Furthermore, if we naively want to relate correlators of the stress-energy tensor to current correlators using \eqref{eq-sugawara}, we would be relating $n$-point correlators of the stress-energy tensor to $2n$-point current correlators in a specific kinematic limit where $n$ pairs of positions coincide. In contrast, the double-copy procedure presented here is fairly straightforward at the level of correlators, and applies directly to the connected part of the correlators. Indeed, it allows us to deduce the $n$-point correlators of the stress-energy tensor from the $n$-point current correlators rather than the $2n$-point function.

\vskip 4pt

To illustrate how to relate stress-energy tensor and current correlators from the Sugawara construction, we obtain in Appendix~\ref{sec-sugawara} the two-point function of the stress-energy tensor from the four-point function of conserved currents as a simple example. Notice that, to match our double-copy prescription to the Sugawara construction, we should further fix the central charge of the stress-energy tensor correlators to be $c=2\hs\gamma\hs k\hs \dim(g)$, which we derive in the same appendix.

\section{Conclusions and Outlook} \label{sec:Conclusions}

We showed that spinor variables in embedding space, which are similar to spinor helicity variables of four-dimensional flat space, are useful in describing particles of any mass and spin in three-dimensional Anti-de Sitter space. Focusing on massless particles is fruitful as their correlation functions have the additional property of holomorphicity, further constraining their dynamical properties. In particular, the dependence on kinematics is fixed at tree level from the structure of the three-point function. This is made manifest in the connection between BCFW recursion relations and BPZ Ward identities of 2d CFT.

\vskip 4pt

This work focused on a very strongly protected sector of the theory. We list a few possibilities for further investigation below: \begin{itemize}
    \item An obvious direction would be to add various levels of supersymmetry to our construction.
    \item It would be interesting to explore less constrained sectors with richer dynamics, like the couplings of massive charged particles to conserved currents. A natural guess is that there is a connection of the bulk dynamics to Knizhnik-Zamolodchikov (KZ) equations, where the correlation functions are expressible in terms of hypergeometric functions.

    \item Another interesting direction would be to consider higher spin currents. Massless theories with higher spin symmetry are often forbidden or tightly constrained; see e.g.~\cite{Henneaux:2010xg, Campoleoni:2010zq, Gaberdiel:2010pz, deBoer:2014fra}. However, we know of algebras with higher spin currents in 2d CFT. Rediscovering them from the bulk using on-shell methods would also be illuminating.
    \item In celestial holography, seemingly related connections between the spacetimes discussed here appear: four-dimensional flat space, three-dimensional Anti-de Sitter space in the form of ``leaf amplitudes", and the two-dimensional CFT on the celestial sphere~\cite{deBoer:2003vf, Fan:2020xjj,Casali:2022fro,Iacobacci:2022yjo,Sleight:2023ojm,Melton:2023bjw,Melton:2024akx,Melton:2024jyq}. Perhaps some of the technology we discussed here can be imported to that context.
\end{itemize} 
The connection between observables in flat and curved spacetime is still in its early stages of exploration. We have shown a simple example of this synergy, connecting modern amplitude techniques to classic results in conformal field theory. Much more is likely still to be uncovered, and we look forward to seeing it.

 \vspace{0.2cm}
 \paragraph{Acknowledgments} We have benefited greatly from the feedback and advice of many colleagues. GLP presented some of these results at a satellite workshop of the Simons Collaboration on celestial holography, and is grateful for questions and feedback from various participants. Special thanks to Daniel Baumann for initial collaboration in this and related projects, and to Jan de Boer, Thomas Dumitrescu and Jo\~ao Penedones for insightful discussions.

\vskip 4pt

GM is supported by the Simons Foundation grant 488649 (Simons Collaboration on the Nonperturbative Bootstrap) and the Swiss National Science Foundation through the project
200020\_197160 and through the National Centre of Competence in Research SwissMAP. GLP and FR are supported by the ERC (NOTIMEFORCOSMO, 101126304), by Scuola Normale, and by INFN (IS GSS-Pi). GLP is also supported by the Italian Ministry of Universities and Research (MUR) under contract 20223ANFHR (PRIN2022). This work was performed in part at Aspen Center for Physics, which is supported by National Science Foundation grant PHY-2210452 and by a grant from the Simons Foundation (1161654, Troyer).
 
\newpage
\appendix
\section{Review of Recursion Relations for Amplitudes} \label{sec:amplitudes-recursion}
In this appendix, we briefly review on-shell recursion relations for scattering amplitudes in four-dimensional Minkowski spacetime following \cite{Cheung:2017pzi,Elvang:2015rqa}. This is useful because the recursion relations for boundary correlators in AdS$_3$ that we derived in this work closely mimic these known results. The fact that they are on-shell makes them conceptually very compelling, because there is no reference to the off-shell ``bulk" spacetime associated with the action formalism and the corresponding Feynman diagram expansion. In what follows, we will study these on-shell recursion relations for the case of tree-level scattering amplitudes of massless particles in four-dimensional Minkowski space.

\subsection{Complex Deformations}

The basic idea of these recursion relations is fairly simple. Ultimately, it boils down to a systematic procedure that allows us to probe the amplitude at singular kinematics. The on-shell kinematics of an amplitude $A$ can be probed by performing a complex deformation of the external momenta
\be
p_i\mapsto p_i(z) = p_i + zq_i\, ,\label{eq:AmpCompDef}
\ee 
for some complex number $z$ that parametrizes the shift that is given by the choice of the vectors $q_i$. Note that, generically, not every momentum needs to be shifted. This shift will induce a deformation of the scattering amplitude as 
\be 
A\mapsto A(z)\, ,
\ee 
and it is clear that $A(z=0)=A$ by construction. Given that we only want to probe on-shell kinematics, the deformed momenta must stay on-shell for any value of the deformation parameter~$z$. This puts constraints on the allowed shift vectors $q_i$:
\be \label{eq:BCFWconstraints}
p_i^2(z )\overset{!}{=} 0\qquad \Longleftrightarrow\qquad  q_i^2 = q_i\cdot p_i=0\, .
\ee 
Moreover, total momentum must remain conserved, which implies 
\be 
\sum_i p_i(z) \overset{!}{=}0\, ,\qquad \Longleftrightarrow \qquad \sum_i q_i = 0\, ,
\ee 
such that the deformed amplitude $A(z)$ remains physical.

\vskip 4pt

We stress that the complex deformation \eqref{eq:AmpCompDef} probes the complexified on-shell kinematics where the momenta can take any complex values and are not restricted by any reality condition. There is thus an implicit step here given by analytically continuing the scattering amplitude to complex momenta. As we will soon see, this will allow us to use the full power of complex analysis and Cauchy's theorem to deduce the undeformed amplitude in terms of residues of the deformed amplitudes under certain assumptions. This deformed amplitude will be given by lower-point amplitudes due to factorization theorems. 

\subsection{Residues from Factorization}

At tree level, the amplitude is a rational function of the momenta. Hence, its analytical continuation to complex momenta is trivial, and the deformed amplitude $A(z)$ is a meromorphic function of the complex parameter $z$. We can use this to write the undeformed amplitude using Cauchy's theorem as 
\be 
A =A(z=0) = \oint_{z=0} \frac{dz}{2\pi i}\frac{A(z)}{z} = -\sum_I \text{Res}_{z=z_I}\left(\frac{A(z)}{z}\right) + B_\infty\, ,\label{eq:Az0}
\ee 
where we expressed the residue of $A(z)/z$ at $z=0$ in terms of the residues at all other poles $z=z_I$, and the residue $-B_\infty$ at $z=\infty$ which represents the boundary contribution from the integral over the circle at infinite radius. 

\vskip 4pt

Physically, this corresponds to the fact that the poles at $z_I$, which label kinematic singularities of the amplitude, necessarily coincide with the appropriate factorization channel of the amplitude. In particular, we know that locality implies that the only possible singularities of the (analytically continued) tree-level amplitude arise when the propagator of an internal particle with momentum $p_I = \sum_{i\in I}p_i$ goes on-shell, where $I$ is a certain subset of external legs. This implies that the deformed amplitude $A(z)$ only has singularities when a deformed internal momentum $p_I(z_I) = p_I + z_I q_I$ goes on-shell
\be 
p_I^2(z_I) = p_I^2 + 2 z_I p_I\cdot q_I = 0 \Leftrightarrow z_I = -\frac{p_I^2 }{2p_I\cdot q_I}\, ,
\ee 
where we assumed for simplicity that $q_i\cdot q_j=0$, thus implying $q_I^2 =0$ for any subset $I$ of external legs. This assumption implies that $p_I^2(z)$ is linear in $z$. Another consequence of locality is that the singularity of the amplitude in $p_I^2(z_I)$ is a simple pole, and that its residue factorizes into lower-point amplitudes as 
\be 
\lim_{z\to z_I} p_I^2(z) A(z) = A_L(z_I)A_R(z_I)\, ,\label{eq:Afactorization}
\ee 
where $A_L(z_I)$ involves the external particles with momenta in the subset $I$ and the internal particle with on-shell momentum $p_I(z_I)$, while $A_R(z_I)$ involves this internal particle together with the remaining external particles. We can thus obtain the residue on the right-hand side of \eqref{eq:Az0} as 
\be 
-\text{Res}_{z=z_I}\left(\frac{A(z)}{z}\right) = -\lim_{z\to z_I}\frac{z-z_I}{z}A(z) = A_L(z_I)\frac{1}{p_I^2}A_R(z_I)\, .
\ee 
We can thus write the amplitude in \eqref{eq:Az0} as a recursion relation in terms of lower-point amplitudes
\be 
A = A(z=0) = \sum_IA_L(z_I)\frac{1}{p_I^2}A_R(z_I) + B_\infty\, ,\label{eq:Arecursion}
\ee 
where $I$ runs over all factorization channels that have shifted momenta on both sides of the channel.\footnote{These correspond to subsets of external legs $I$, where both $I$ and its complement $\bar{I}$ have
shifted momenta. Indeed, if there were no shifts in $\bar{I}$, momentum conservation would keep
$p_I(z)= -p_{\bar{I}}$ constant, and thus no singularity would appear.}

\vskip 4pt

In cases where the residue at infinity vanishes (i.e.~$B_\infty=0$) for some momentum shift, the theory is said to be ``on-shell reconstructible" because the recursion relations in~\eqref{eq:Arecursion} allow us to deduce any amplitude from lower-point amplitudes in a way that is purely on-shell. Conveniently, this criterion is satisfied for Yang--Mills theory, gravity, and for any renormalizable quantum field theory in four-dimensional Minkowski space \cite{Cheung:2008dn,Cohen:2010mi,Cheung:2017pzi}, such as the whole Standard Model.

\vskip 4pt

While it is straightforward to bootstrap the three-point scattering amplitude using only the restrictions coming from kinematics~\cite{Benincasa:2007xk,Arkani-Hamed:2017jhn,Cheung:2017pzi} (that is, Poincar\'e invariance, little group covariance, and gauge invariance), these constraints are not enough to determine higher-point amplitudes. It is clear that some dynamical input is needed to determine the latter. We have just learned that by assuming locality and a vanishing boundary contribution, we can deduce all higher-point amplitudes at tree level by using the recursion relations in \eqref{eq:Arecursion}. Thus, the extra input coming from dynamics is that the interactions are local, which implied the factorization property \eqref{eq:Afactorization} of the amplitudes, and that the theory is on-shell reconstructible so that there is no boundary contribution in \eqref{eq:Arecursion} (it is sufficient to assume that the theory is renormalizable).

\subsection{BCFW Recursion}

As we have learned in the introduction of Section \ref{subsec:spinorhelicity}, the spinor helicity variables parameterize the on-shell kinematics of scattering amplitudes without any gauge redundancy, and they make manifest that each external momentum is on-shell. Hence, they will be useful to trivialize the constraint \eqref{eq:BCFWconstraints} that the deformed momenta must remain on-shell. Indeed, it is straightforward to show that any deformation of the form \eqref{eq:AmpCompDef} that satisfies this constraint must be of one of the two following forms:
\be \label{eq-generic-deform-amplitudes}
\begin{aligned} 
p_{i,\alpha\dot\alpha}(z)&=\lambda_{i,\alpha}\tilde\lambda_{i,\dot\alpha}(z)\,,\quad\text{with}\quad\tilde\lambda_{i,\dot\alpha}(z)=\tilde\lambda_{i,\dot\alpha}+z\hs \tilde\eta_{i,\dot\alpha}\,,\\
p_{i,\alpha\dot\alpha}(z)&=\lambda_{i,\alpha}(z)\tilde\lambda_{i,\dot\alpha}\,,\quad\text{with}\quad \lambda_{i,\alpha}(z)=\lambda_{i,\alpha}+z\hs \eta_{i,\alpha}\,,
\end{aligned}
\ee 
where we wrote the momenta $p_{i,\alpha\dot\alpha}=p_{i,\mu}(\sigma^\mu)_{\alpha\dot\alpha}$ as a $2\times 2$ matrix, which must have a vanishing determinant for the momenta to be on-shell. In fact, we are trivializing this by writing the matrix as an outer product of two spinors. Here, the precise complex deformation is given by the choice of the spinors $\eta_i$ or $\tilde \eta_i$. However, this choice must be consistent with momentum conservation, which yields another constraint that is not trivialized.

\vskip 4pt

Let us consider a very simple and important complex deformation, the so-called BCFW shift~\cite{Britto:2004ap,Britto:2005fq} given by shifting only the angle and square spinors of the two external legs $1$ and $2$, respectively, as 
\begin{align}
\tilde\lambda_1&\mapsto \tilde\lambda_1(z)=\tilde\lambda_1+z\tilde\lambda_2\,,\\
\lambda_2&\mapsto \lambda_2(z)=\lambda_2-z\lambda_1\,,
\end{align}
which automatically preserves both the on-shell constraint \eqref{eq:BCFWconstraints} as it has the form of \eqref{eq-generic-deform-amplitudes}, and the constraint coming from momentum conservation because 
\be 
p_1(z)+p_2(z)=(p_1+z \hs\lambda_1\tilde \lambda_2)+(p_2-z\hs \lambda_1\tilde \lambda_2)=p_1+p_2\,.
\ee 

Remarkably, the boundary contribution $B_\infty$ vanishes under this complex deformation for both Yang--Mills theory and gravity as long as the momentum shift is applied to the external legs $1,2$ with a helicity configuration that is not $1^+2^-$ \cite{Benincasa:2007qj,Arkani-Hamed:2008bsc, Cheung:2017pzi}. Hence, we can always choose a pair of gluons for the BCFW shift so that there is no boundary contribution, and thus the theory is on-shell reconstructible.

\vskip 4pt

Note that we can therefore write any higher-point amplitude in YM theory or gravity in terms of the three-point amplitude, which is fixed by kinematics. Hence, the four-point vertex of Yang--Mills theory and the infinite tower of higher-order graviton vertices have no physical content beyond the cubic interaction vertex. Their only purpose is to maintain gauge or diffeomorphism invariance for YM or GR, respectively, at the level of the action.

\vskip 4pt 

Notably, this recursion can be used to derive the famous Parke--Taylor formula \cite{Parke:1986gb,Berends:1987me} for the color-ordered scattering amplitude of gluons in YM theory in the maximally helicity-violating (MHV) configuration:
\be \label{eq-Parke--Taylor-amplitudes}
A_n(1^-2^-3^+\cdots n^+)=\frac{\langle 12 \rangle^4}{\langle 12 \rangle \langle 23 \rangle\cdots \langle n1\rangle}\,,
\ee 
where exactly two particles have helicity~$-$ and all the others have helicity $+$. A detailed proof of this can be found in Section~3.2 of \cite{Elvang:2015rqa}.

\section{Sugawara Construction}\label{sec-sugawara}

In this appendix, we show how to relate current correlators to stress-energy tensor correlators obtained via the Sugawara construction. In particular, we will evaluate the current four-point function, including its disconnected terms, and take the appropriate coincident limit to see this as a Sugawara stress-energy tensor two-point function. This calculation will allow us to derive the correct central charge and check consistency. 

\vskip 4pt

Consider a holomorphic affine current algebra. For two such currents, the OPE in position space is simply \eqref{equ:OPE-JJ}, which we recall here for convenience 
\begin{equation}
J^a(w_1)J^b(w_2)
\sim
\frac{k\,\delta^{ab}}{(w_1-w_2)^2}
+
\frac{i f^{ab}{}_{c}\,J^c(w_2)}{w_1-w_2}\, ,
\end{equation}
with level $k$ and structure constants $f^{abc}$. The Ward identity following from the current symmetry is
\begin{align}
&\braket{J^a(w)J^{a_1}(w_1)\dots J^{a_n}(w_n)} =\label{eq:WIapp} \\
&\quad\quad     \sum_{i=1}^n\Big[\frac{k\delta^{aa_i}}{(w-w_i)^2}\braket{J^{a_1}\dots J^{a_{i-1}}J^{a_{i+1}}\dots J^{a_n}}+\frac{if^{aa_i}{}_{c}}{w-w_i}\braket{J^{a_1}\dots J^{a_{i-1}} J^c(w_i)J^{a_{i+1}}\dots J^{a_n}}\Big]\, ,\nonumber
\end{align}
where in the first term, the operator $J^{a_i}(w_i)$ is removed from the correlators and in the second term $J^{a_i}(w_i)$ is replaced by $J^c(w_i)$. Starting from 
\begin{equation}
    \braket{J^a(w_1)J^b(w_2)} = \frac{k\delta^{ab}}{w_{12}^2}\, ,\label{eq:JJexp}
\end{equation}
with $w_{ij}\equiv w_i-w_j$, we can use \eqref{eq:WIapp} to derive the three-point function
\begin{align}
\braket{J^a(w)J^{a_1}(w_1)J^{a_2}(w_2)} &=  \frac{if^{aa_1}{}_{d}}{(w-w_1)}\braket{J^d(w_1)J^{a_2}(w_2)}+\frac{if^{aa_2}{}_{d}}{(w-w_2)}\braket{J^{a_1}(w_1)J^d(w_2)} \nonumber\\ 
&= \frac{ik f^{aa_1a_2}}{w_{12}(w-w_1)(w-w_2)}\, .
\end{align}
Note that the contributions from the double pole vanish because the one-point function is zero. The three-point function is thus
\begin{equation}
    \braket{J^a_1 J^b_2 J^c_3} = \frac{i k f^{abc}}{w_{12}w_{23}w_{13}}\, .\label{eq:JJJexpl}
\end{equation}
The Sugawara stress-energy tensor is defined as \cite{Sugawara:1968gq}
\begin{equation}
T(w)
=
\frac{1}{2(k+h^\vee)}
:\!J^aJ^a\!:(w),\label{eq:Sugawaradef}
\end{equation}
where the dual Coxeter number is defined by
\begin{equation}
f^{a}{}_{cd} f^{bcd}
=
2h^\vee \delta^{ab},\label{eq:coxeter}
\end{equation}
and the two currents are normal ordered. Our goal is to compute the stress-energy tensor two-point function using the definition \eqref{eq:Sugawaradef} to be able to read off the central charge $c$ as it is generically encoded in this correlator as 
\begin{equation}
\langle T(w_1)T(w_3)\rangle
=
\frac{c/2}{w_{13}^4}.\label{eq:2ptTapp}
\end{equation}
Substituting the Sugawara definition \eqref{eq:Sugawaradef}, the two-point function is then related to the current four-point function as 
\begin{equation}
\langle T(w_1)T(w_3)\rangle
=
\frac{1}{4(k+h^\vee)^2}
\,
\Big\langle
:\!J^aJ^a\!:(w_1)
:\!J^bJ^b\!:(w_3)
\Big\rangle .\label{eq:TTSugawara}
\end{equation}
We can build the relevant four-current correlator using the Ward identity \eqref{eq:WIapp} together with the explicit result for the two- and three-point functions \eqref{eq:JJexp} and \eqref{eq:JJJexpl}. We obtain
\begin{align}
\langle
J^{a}(w_1)J^{b}(w_2)J^{c}(w_3)J^{d}(w_4)
\rangle
&=
k^2\Bigg[
\frac{\delta^{ab}\delta^{cd}}{w_{12}^2 w_{34}^2}
+\frac{\delta^{ac}\delta^{bd}}{w_{13}^2 w_{24}^2}
+\frac{\delta^{ad}\delta^{bc}}{w_{14}^2 w_{23}^2}
\Bigg]
\label{eqJJJJap}\\
&\quad
-k\Bigg[
\frac{f^{ab}{}_e f^{cde}}
     {w_{12}w_{23}w_{34}w_{24}}
+\frac{f^{ac}{}_e f^{dbe}}
     {w_{13}w_{23}w_{34}w_{24}}
+\frac{f^{ad}{}_e f^{bce}}
     {w_{14}w_{23}w_{34}w_{24}}
\Bigg].\nonumber 
\end{align}
To obtain the correlator of the composite operators as in \eqref{eq:TTSugawara}, we consider the current four-point function $
\langle
J^a(w_1)J^a(w_2)
J^b(w_3)J^b(w_4)
\rangle
$
and take the coincident-point limit:
\begin{equation}
w_2 \to w_1\,,
\qquad\text{and}\qquad 
w_4 \to w_3\,,
\end{equation}
while subtracting the singularities internal to each normal-ordered pair. Taking the coincident limit for the terms in \eqref{eqJJJJap} that are proportional to $k^2$ gives
\begin{equation}
\lim_{\substack{w_{2}\rightarrow w_1\\
w_{4}\rightarrow w_3}}\langle
J^{a}(w_1)J^{a}(w_2)J^{b}(w_3)J^{b}(w_4)\rangle\Big|_{\mathcal{O}(k^2)}
\supset\frac{2k^2\,\dim(g)}{(w_1-w_3)^4} ,
\end{equation}
where $\dim(g)$ is the dimension of the Lie algebra, and we discarded the first term on the right-hand side of \eqref{eqJJJJap} because it is divergent in this limit. For the terms that are proportional to $k$ in \eqref{eqJJJJap}, we get 
\begin{align}
\lim_{\substack{w_{2}\rightarrow w_1\\
w_{4}\rightarrow w_3}}\langle
J^{a}(w_1)J^{a}(w_2)J^{b}(w_3)J^{b}(w_4)\rangle\Big|_{\mathcal{O}(k)}
&\supset-kf^{ab}{}_e f^{abe}\Bigg[
\frac{(-1)}
     {(w_1-w_3)^4}
\Bigg]=
\frac{2kh^\vee \dim(g)}
     {(w_1-w_3)^4}
\, ,
\end{align}
where the first term vanishes identically because $f\indices{^{aa}_e}=0$, while the individual $w_{34}^{-1}$ divergences cancel between the two remaining structures. In the last step we used \eqref{eq:coxeter}.
Collecting the two terms, we have the following 
\begin{align}
\langle
:J^{a}J^{a}:(w_1)\,:J^{b}J^{b}:(w_3)
\rangle
&= 2k\dim(g)(k + h^\vee)\frac{1}{(w_1-w_3)^4}\, .\label{eq:B19}
\end{align}
Comparing the Sugawara expression  \eqref{eq:TTSugawara}, evaluated using \eqref{eq:B19}, with the standard normalization \eqref{eq:2ptTapp}, we obtain
\begin{equation}
    \frac{c}{2} = \frac{2k\dim (g)(k+h^\vee)}{4(k+h^\vee)^2} = \frac{k\dim(g)}{2(k+h^\vee)}\, ,
\end{equation}
such that the central charge is just 
\begin{equation}
c
=
\frac{k\,\dim(g)}{k+h^\vee}.
\end{equation}
This is the correct answer for the Sugawara construction, which we recover from consistency of current correlators.

\newpage
\phantomsection
\addcontentsline{toc}{section}{References}
\bibliographystyle{utphys}
{\linespread{1.075}
	\bibliography{SHV-CFT2-Refs}

@article{Britto:2005fq,
	title        = {{Direct Proof of Tree-Level Recursion Relation in Yang-Mills Theory}},
	author       = {Britto, Ruth and Cachazo, Freddy and Feng, Bo and Witten, Edward},
	year         = 2005,
	journal      = {Phys. Rev. Lett.},
	volume       = 94,
	pages        = 181602,
	doi          = {10.1103/PhysRevLett.94.181602},
	archiveprefix = {arXiv},
	eprint       = {hep-th/0501052}
}

@inproceedings{Ginsparg:1988ui,
	title        = {{Applied Conformal Field Theory}},
	author       = {Ginsparg, Paul H.},
	year         = 1988,
	month        = 9,
	booktitle    = {{Les Houches Summer School in Theoretical Physics: Fields, Strings, Critical Phenomena}},
	archiveprefix = {arXiv},
	eprint       = {hep-th/9108028},
	reportnumber = {HUTP-88-A054}
}

@article{Moore:1989yh,
	title        = {{Taming the Conformal Zoo}},
	author       = {Moore, Gregory W. and Seiberg, Nathan},
	year         = 1989,
	journal      = {Phys. Lett. B},
	volume       = 220,
	pages        = {422--430},
	doi          = {10.1016/0370-2693(89)90897-6},
	reportnumber = {IASSNS-HEP-89/6}
}

@article{Elitzur:1989nr,
	title        = {{Remarks on the Canonical Quantization of the Chern-Simons-Witten Theory}},
	author       = {Elitzur, Shmuel and Moore, Gregory W. and Schwimmer, Adam and Seiberg, Nathan},
	year         = 1989,
	journal      = {Nucl. Phys. B},
	volume       = 326,
	pages        = {108--134},
	doi          = {10.1016/0550-3213(89)90436-7},
	reportnumber = {IASSNS-HEP-89/20}
}

@article{Witten:1988hf,
	title        = {{Quantum Field Theory and the Jones Polynomial}},
	author       = {Witten, Edward},
	year         = 1989,
	journal      = {Commun. Math. Phys.},
	volume       = 121,
	pages        = {351--399},
	doi          = {10.1007/BF01217730},
	editor       = {Mitra, Asoke N.},
	reportnumber = {IASSNS-HEP-88-33}
}

@article{Sugawara:1968gq,
	title        = {{A Field Theory of Currents}},
	author       = {Sugawara, Hirotaka},
	year         = 1968,
	journal      = {Phys. Rev.},
	volume       = 170,
	pages        = {1659--1662},
	doi          = {10.1103/PhysRev.170.1659}
}

@article{Fan:2020xjj,
	title        = {{On Sugawara construction on Celestial Sphere}},
	author       = {Fan, Wei and Fotopoulos, Angelos and Stieberger, Stephan and Taylor, Tomasz R.},
	year         = 2020,
	journal      = {JHEP},
	volume       = {09},
	pages        = 139,
	doi          = {10.1007/JHEP09(2020)139},
	eprint       = {2005.10666},
	archiveprefix = {arXiv},
	primaryclass = {hep-th}
}

@article{BELAVIN1984333,
	title        = {Infinite conformal symmetry in two-dimensional quantum field theory},
	author       = {A.A. Belavin and A.M. Polyakov and A.B. Zamolodchikov},
	year         = 1984,
	journal      = {Nuclear Physics B},
	volume       = 241,
	number       = 2,
	pages        = {333--380},
	doi          = {https://doi.org/10.1016/0550-3213(84)90052-X},
	issn         = {0550-3213},
	url          = {https://www.sciencedirect.com/science/article/pii/055032138490052X}
}

@article{deBoer:2014fra,
	title        = {{Boundary conditions and partition functions in higher spin AdS$_{3}$/CFT$_{2}$}},
	author       = {de Boer, Jan and Jottar, Juan I.},
	year         = 2016,
	journal      = {JHEP},
	volume       = {04},
	pages        = 107,
	doi          = {10.1007/JHEP04(2016)107},
	eprint       = {1407.3844},
	archiveprefix = {arXiv},
	primaryclass = {hep-th}
}

@article{Bhattacharya:2025udq,
	title        = {{Chern-Simons propagators in AdS$_3$}},
	author       = {Bhattacharya, Jyotirmoy and Guria, Anurag and Prakash, Shiroman and Sharma, Aditya and Sharma, Tarun},
	year         = 2025,
	month        = 12,
	eprint       = {2512.07752},
	archiveprefix = {arXiv},
	primaryclass = {hep-th}
}

@article{Keranen:2014ava,
	title        = {{Chern-Simons interactions in AdS$_3$ and the current conformal block}},
	author       = {Keranen, Ville},
	year         = 2014,
	month        = 3,
	eprint       = {1403.6881},
	archiveprefix = {arXiv},
	primaryclass = {hep-th}
}

@article{Witten:1998wy,
	title        = {{AdS/CFT correspondence and topological field theory.}},
	author       = {Witten, Edward},
	year         = 1998,
	journal      = {JHEP},
	volume       = 12,
	pages        = {012},
	doi          = {10.1088/1126-6708/1998/12/012},
	eprint       = {hep-th/9812012},
	archiveprefix = {arXiv},
	reportnumber = {IASSNS-HEP-98-96}
}

@inproceedings{Gukov:2004id,
	title        = {{Chern-Simons gauge theory and the AdS(3) / CFT(2) correspondence}},
	author       = {Gukov, Sergei and Martinec, Emil and Moore, Gregory W. and Strominger, Andrew},
	year         = 2004,
	month        = 3,
	booktitle    = {{From Fields to Strings: Circumnavigating Theoretical Physics: A Conference in Tribute to Ian Kogan}},
	pages        = {1606--1647},
	doi          = {10.1142/9789812775344_0036},
	eprint       = {hep-th/0403225},
	archiveprefix = {arXiv},
	reportnumber = {HUTP-04-A014}
}

@article{Raju:2010by,
	title        = {{BCFW for Witten Diagrams}},
	author       = {Raju, Suvrat},
	year         = 2011,
	journal      = {Phys. Rev. Lett.},
	volume       = 106,
	pages        = {091601},
	doi          = {10.1103/PhysRevLett.106.091601},
	eprint       = {1011.0780},
	archiveprefix = {arXiv},
	primaryclass = {hep-th},
	reportnumber = {HRI-ST-1009}
}

@article{Zhou:2018sfz,
	title        = {{Recursion Relations in Witten Diagrams and Conformal Partial Waves}},
	author       = {Zhou, Xinan},
	year         = 2019,
	journal      = {JHEP},
	volume       = {05},
	pages        = {006},
	doi          = {10.1007/JHEP05(2019)006},
	eprint       = {1812.01006},
	archiveprefix = {arXiv},
	primaryclass = {hep-th},
	reportnumber = {PUPT-2575}
}

@article{Diwakar:2021juk,
	title        = {{BCJ amplitude relations for Anti-de Sitter boundary correlators in embedding space}},
	author       = {Diwakar, Pranav and Herderschee, Aidan and Roiban, Radu and Teng, Fei},
	year         = 2021,
	journal      = {JHEP},
	volume       = 10,
	pages        = 141,
	doi          = {10.1007/JHEP10(2021)141},
	eprint       = {2106.10822},
	archiveprefix = {arXiv},
	primaryclass = {hep-th},
	reportnumber = {LCTP-21-15}
}

@article{Albayrak:2020fyp,
	title        = {{On duality of color and kinematics in (A)dS momentum space}},
	author       = {Albayrak, Soner and Kharel, Savan and Meltzer, David},
	year         = 2021,
	journal      = {JHEP},
	volume       = {03},
	pages        = 249,
	doi          = {10.1007/JHEP03(2021)249},
	eprint       = {2012.10460},
	archiveprefix = {arXiv},
	primaryclass = {hep-th}
}

@article{Zhou:2021gnu,
	title        = {{Double Copy Relation in AdS Space}},
	author       = {Zhou, Xinan},
	year         = 2021,
	journal      = {Phys. Rev. Lett.},
	volume       = 127,
	number       = 14,
	pages        = 141601,
	doi          = {10.1103/PhysRevLett.127.141601},
	eprint       = {2106.07651},
	archiveprefix = {arXiv},
	primaryclass = {hep-th}
}

@article{Lipstein:2019mpu,
	title        = {{Double copy structure and the flat space limit of conformal correlators in even dimensions}},
	author       = {Lipstein, Arthur E. and McFadden, Paul},
	year         = 2020,
	journal      = {Phys. Rev. D},
	volume       = 101,
	number       = 12,
	pages        = 125006,
	doi          = {10.1103/PhysRevD.101.125006},
	eprint       = {1912.10046},
	archiveprefix = {arXiv},
	primaryclass = {hep-th}
}

@article{Witten:1988hc,
	title        = {{(2+1)-Dimensional Gravity as an Exactly Soluble System}},
	author       = {Witten, Edward},
	year         = 1988,
	journal      = {Nucl. Phys. B},
	volume       = 311,
	pages        = 46,
	doi          = {10.1016/0550-3213(88)90143-5},
	reportnumber = {IASSNS-HEP-88-32}
}

@article{Feng:2009ei,
	title        = {{BCFW Recursion Relation with Nonzero Boundary Contribution}},
	author       = {Feng, Bo and Wang, Junqi and Wang, Yihong and Zhang, Zhibai},
	year         = 2010,
	journal      = {JHEP},
	volume       = {01},
	pages        = {019},
	doi          = {10.1007/JHEP01(2010)019},
	eprint       = {0911.0301},
	archiveprefix = {arXiv},
	primaryclass = {hep-th}
}

@article{Cheung:2008dn,
	title        = {{On-Shell Recursion Relations for Generic Theories}},
	author       = {Cheung, Clifford},
	year         = 2010,
	journal      = {JHEP},
	volume       = {03},
	pages        = {098},
	doi          = {10.1007/JHEP03(2010)098},
	eprint       = {0808.0504},
	archiveprefix = {arXiv},
	primaryclass = {hep-th}
}

@article{Cohen:2010mi,
	title        = {{On-shell constructibility of tree amplitudes in general field theories}},
	author       = {Cohen, Timothy and Elvang, Henriette and Kiermaier, Michael},
	year         = 2011,
	journal      = {JHEP},
	volume       = {04},
	pages        = {053},
	doi          = {10.1007/JHEP04(2011)053},
	eprint       = {1010.0257},
	archiveprefix = {arXiv},
	primaryclass = {hep-th},
	reportnumber = {MCTP-10-46, PUPT-2351}
}

@article{Benincasa:2007qj,
	title        = {{Taming Tree Amplitudes In General Relativity}},
	author       = {Benincasa, Paolo and Boucher-Veronneau, Camille and Cachazo, Freddy},
	year         = 2007,
	journal      = {JHEP},
	volume       = 11,
	pages        = {057},
	doi          = {10.1088/1126-6708/2007/11/057},
	eprint       = {hep-th/0702032},
	archiveprefix = {arXiv}
}

@article{Arkani-Hamed:2008bsc,
	title        = {{On Tree Amplitudes in Gauge Theory and Gravity}},
	author       = {Arkani-Hamed, Nima and Kaplan, Jared},
	year         = 2008,
	journal      = {JHEP},
	volume       = {04},
	pages        = {076},
	doi          = {10.1088/1126-6708/2008/04/076},
	eprint       = {0801.2385},
	archiveprefix = {arXiv},
	primaryclass = {hep-th}
}

@article{Arkani-Hamed:2017jhn,
	title        = {{Scattering amplitudes for all masses and spins}},
	author       = {Arkani-Hamed, Nima and Huang, Tzu-Chen and Huang, Yu-tin},
	year         = 2021,
	journal      = {JHEP},
	volume       = 11,
	pages        = {070},
	doi          = {10.1007/JHEP11(2021)070},
	eprint       = {1709.04891},
	archiveprefix = {arXiv},
	primaryclass = {hep-th},
	reportnumber = {NCTS-TH/1714, NCTS-TH-1714}
}

@article{Benincasa:2007xk,
	title        = {{Consistency Conditions on the S-Matrix of Massless Particles}},
	author       = {Benincasa, Paolo and Cachazo, Freddy},
	year         = 2007,
	month        = 5,
	eprint       = {0705.4305},
	archiveprefix = {arXiv},
	primaryclass = {hep-th},
	reportnumber = {UWO-TH-07-09}
}

@article{Berends:1987me,
	title        = {{Recursive Calculations for Processes with n Gluons}},
	author       = {Berends, Frits A. and Giele, W. T.},
	year         = 1988,
	journal      = {Nucl. Phys. B},
	volume       = 306,
	pages        = {759--808},
	doi          = {10.1016/0550-3213(88)90442-7},
	reportnumber = {Print-88-0100 (LEIDEN)}
}

@article{Parke:1986gb,
	title        = {{An Amplitude for $n$ Gluon Scattering}},
	author       = {Parke, Stephen J. and Taylor, T. R.},
	year         = 1986,
	journal      = {Phys. Rev. Lett.},
	volume       = 56,
	pages        = 2459,
	doi          = {10.1103/PhysRevLett.56.2459},
	reportnumber = {FERMILAB-PUB-86-042-T}
}

@article{Gaberdiel:2010pz,
	title        = {{An AdS$_{3}$ Dual for Minimal Model CFTs}},
	author       = {Gaberdiel, Matthias R. and Gopakumar, Rajesh},
	year         = 2011,
	journal      = {Phys. Rev. D},
	volume       = 83,
	pages        = {066007},
	doi          = {10.1103/PhysRevD.83.066007},
	eprint       = {1011.2986},
	archiveprefix = {arXiv},
	primaryclass = {hep-th}
}

@article{Campoleoni:2010zq,
	title        = {{Asymptotic symmetries of three-dimensional gravity coupled to higher-spin fields}},
	author       = {Campoleoni, Andrea and Fredenhagen, Stefan and Pfenninger, Stefan and Theisen, Stefan},
	year         = 2010,
	journal      = {JHEP},
	volume       = 11,
	pages        = {007},
	doi          = {10.1007/JHEP11(2010)007},
	eprint       = {1008.4744},
	archiveprefix = {arXiv},
	primaryclass = {hep-th},
	reportnumber = {AEI-2010-140}
}

@article{Henneaux:2010xg,
	title        = {{Nonlinear $W_{infinity}$ as Asymptotic Symmetry of Three-Dimensional Higher Spin Anti-de Sitter Gravity}},
	author       = {Henneaux, Marc and Rey, Soo-Jong},
	year         = 2010,
	journal      = {JHEP},
	volume       = 12,
	pages        = {007},
	doi          = {10.1007/JHEP12(2010)007},
	eprint       = {1008.4579},
	archiveprefix = {arXiv},
	primaryclass = {hep-th}
}

@article{Melton:2024akx,
	title        = {{Celestial Dual for Maximal Helicity Violating Amplitudes}},
	author       = {Melton, Walker and Sharma, Atul and Strominger, Andrew and Wang, Tianli},
	year         = 2024,
	journal      = {Phys. Rev. Lett.},
	volume       = 133,
	number       = 9,
	pages        = {091603},
	doi          = {10.1103/PhysRevLett.133.091603},
	eprint       = {2403.18896},
	archiveprefix = {arXiv},
	primaryclass = {hep-th}
}

@article{deBoer:2003vf,
	title        = {{A Holographic reduction of Minkowski space-time}},
	author       = {de Boer, Jan and Solodukhin, Sergey N.},
	year         = 2003,
	journal      = {Nucl. Phys. B},
	volume       = 665,
	pages        = {545--593},
	doi          = {10.1016/S0550-3213(03)00494-2},
	eprint       = {hep-th/0303006},
	archiveprefix = {arXiv},
	reportnumber = {ITFA-2003-11}
}

@article{Sleight:2023ojm,
	title        = {{Celestial Holography Revisited}},
	author       = {Sleight, Charlotte and Taronna, Massimo},
	year         = 2024,
	journal      = {Phys. Rev. Lett.},
	volume       = 133,
	number       = 24,
	pages        = 241601,
	doi          = {10.1103/PhysRevLett.133.241601},
	eprint       = {2301.01810},
	archiveprefix = {arXiv},
	primaryclass = {hep-th}
}

@article{Iacobacci:2022yjo,
	title        = {{From celestial correlators to AdS, and back}},
	author       = {Iacobacci, Lorenzo and Sleight, Charlotte and Taronna, Massimo},
	year         = 2023,
	journal      = {JHEP},
	volume       = {06},
	pages        = {053},
	doi          = {10.1007/JHEP06(2023)053},
	eprint       = {2208.01629},
	archiveprefix = {arXiv},
	primaryclass = {hep-th}
}

@article{Melton:2024jyq,
	title        = {{Soft algebras for leaf amplitudes}},
	author       = {Melton, Walker and Sharma, Atul and Strominger, Andrew},
	year         = 2024,
	journal      = {JHEP},
	volume       = {07},
	pages        = {070},
	doi          = {10.1007/JHEP07(2024)070},
	eprint       = {2402.04150},
	archiveprefix = {arXiv},
	primaryclass = {hep-th}
}

@article{Melton:2023bjw,
	title        = {{Celestial leaf amplitudes}},
	author       = {Melton, Walker and Sharma, Atul and Strominger, Andrew},
	year         = 2024,
	journal      = {JHEP},
	volume       = {07},
	pages        = 132,
	doi          = {10.1007/JHEP07(2024)132},
	eprint       = {2312.07820},
	archiveprefix = {arXiv},
	primaryclass = {hep-th}
}

@article{Casali:2022fro,
	title        = {{Celestial amplitudes as AdS-Witten diagrams}},
	author       = {Casali, Eduardo and Melton, Walker and Strominger, Andrew},
	year         = 2022,
	journal      = {JHEP},
	volume       = 11,
	pages        = 140,
	doi          = {10.1007/JHEP11(2022)140},
	eprint       = {2204.10249},
	archiveprefix = {arXiv},
	primaryclass = {hep-th}
}

@article{Brown:1986nw,
	title        = {{Central Charges in the Canonical Realization of Asymptotic Symmetries: An Example from Three-Dimensional Gravity}},
	author       = {Brown, J. David and Henneaux, M.},
	year         = 1986,
	journal      = {Commun. Math. Phys.},
	volume       = 104,
	pages        = {207--226},
	doi          = {10.1007/BF01211590}
}

@article{Achucarro:1986uwr,
	title        = {{A Chern-Simons Action for Three-Dimensional anti-De Sitter Supergravity Theories}},
	author       = {Achucarro, A. and Townsend, P. K.},
	year         = 1986,
	journal      = {Phys. Lett. B},
	volume       = 180,
	pages        = 89,
	doi          = {10.1016/0370-2693(86)90140-1},
	editor       = {Salam, A. and Sezgin, E.},
	reportnumber = {Print-87-0078 (CAMBRIDGE)}
}

@article{Skvortsov:2022wzo,
	title        = {{On (spinor)-helicity and bosonization in AdS$_{4}$/CFT$_{3}$}},
	author       = {Skvortsov, Evgeny and Yin, Yihao},
	year         = 2023,
	journal      = {JHEP},
	volume       = {03},
	pages        = 204,
	doi          = {10.1007/JHEP03(2023)204},
	eprint       = {2207.06976},
	archiveprefix = {arXiv},
	primaryclass = {hep-th}
}

@article{Jain:2021vrv,
	title        = {{Higher spin 3-point functions in 3d CFT using spinor-helicity variables}},
	author       = {Jain, Sachin and John, Renjan Rajan and Mehta, Abhishek and Nizami, Amin A. and Suresh, Adithya},
	year         = 2021,
	journal      = {JHEP},
	volume       = {09},
	pages        = {041},
	doi          = {10.1007/JHEP09(2021)041},
	eprint       = {2106.00016},
	archiveprefix = {arXiv},
	primaryclass = {hep-th}
}

@article{CarrilloGonzalez:2026eum,
	title        = {{4d CFT Correlators from Ambitwistors}},
	author       = {Carrillo Gonz{\'a}lez, Mariana and Keseman, Th{\'e}o},
	year         = 2026,
	month        = 6,
	eprint       = {2606.13770},
	archiveprefix = {arXiv},
	primaryclass = {hep-th},
	reportnumber = {Imperial/TP/2026/MC/02}
}

@article{CarrilloGonzalez:2025qjk,
	title        = {{Spinning boundary correlators from (A)dS$_{4}$ twistors}},
	author       = {Carrillo Gonz{\'a}lez, Mariana and Keseman, Th{\'e}o},
	year         = 2026,
	journal      = {JHEP},
	volume       = {03},
	pages        = 131,
	doi          = {10.1007/JHEP03(2026)131},
	eprint       = {2510.00096},
	archiveprefix = {arXiv},
	primaryclass = {hep-th},
	reportnumber = {Imperial/TP/2025/MC/02}
}

@article{Beetar:2024ptv,
	title        = {{Double copy in AdS$_{3}$ from minitwistor space}},
	author       = {Beetar, Cameron and Carrillo Gonz{\'a}lez, Mariana and Jaitly, Sumer and Keseman, Th{\'e}o},
	year         = 2025,
	journal      = {JHEP},
	volume       = {03},
	pages        = 125,
	doi          = {10.1007/JHEP03(2025)125},
	eprint       = {2410.23342},
	archiveprefix = {arXiv},
	primaryclass = {hep-th},
	reportnumber = {Imperial/TP/2024/MC/02}
}

@article{Bern:2010ue,
	title        = {{Perturbative Quantum Gravity as a Double Copy of Gauge Theory}},
	author       = {Bern, Zvi and Carrasco, John Joseph M. and Johansson, Henrik},
	year         = 2010,
	journal      = {Phys. Rev. Lett.},
	volume       = 105,
	pages        = {061602},
	doi          = {10.1103/PhysRevLett.105.061602},
	eprint       = {1004.0476},
	archiveprefix = {arXiv},
	primaryclass = {hep-th},
	reportnumber = {UCLA-10-TEP-102, SACLAY-IPHT-T10-044}
}

@article{Bern:2008qj,
	title        = {{New Relations for Gauge-Theory Amplitudes}},
	author       = {Bern, Z. and Carrasco, J. J. M. and Johansson, Henrik},
	year         = 2008,
	journal      = {Phys. Rev. D},
	volume       = 78,
	pages        = {085011},
	doi          = {10.1103/PhysRevD.78.085011},
	eprint       = {0805.3993},
	archiveprefix = {arXiv},
	primaryclass = {hep-ph},
	reportnumber = {UCLA-07-TEP-15}
}

@article{Lee:2022fgr,
	title        = {{Cosmological double-copy relations}},
	author       = {Lee, Hayden and Wang, Xinkang},
	year         = 2023,
	journal      = {Phys. Rev. D},
	volume       = 108,
	number       = 6,
	pages        = {L061702},
	doi          = {10.1103/PhysRevD.108.L061702},
	eprint       = {2212.11282},
	archiveprefix = {arXiv},
	primaryclass = {hep-th}
}

@article{Cheung:2022pdk,
	title        = {{On-shell correlators and color-kinematics duality in curved symmetric spacetimes}},
	author       = {Cheung, Clifford and Parra-Martinez, Julio and Sivaramakrishnan, Allic},
	year         = 2022,
	journal      = {JHEP},
	volume       = {05},
	pages        = {027},
	doi          = {10.1007/JHEP05(2022)027},
	eprint       = {2201.05147},
	archiveprefix = {arXiv},
	primaryclass = {hep-th},
	reportnumber = {CALT-TH-2022-002}
}

@article{Alday:2021odx,
	title        = {{Gluon Scattering in AdS from CFT}},
	author       = {Alday, Luis F. and Behan, Connor and Ferrero, Pietro and Zhou, Xinan},
	year         = 2021,
	journal      = {JHEP},
	volume       = {06},
	pages        = {020},
	doi          = {10.1007/JHEP06(2021)020},
	eprint       = {2103.15830},
	archiveprefix = {arXiv},
	primaryclass = {hep-th}
}

@article{Armstrong:2020woi,
	title        = {{Color/kinematics duality in AdS$_{4}$}},
	author       = {Armstrong, Connor and Lipstein, Arthur E. and Mei, Jiajie},
	year         = 2021,
	journal      = {JHEP},
	volume       = {02},
	pages        = 194,
	doi          = {10.1007/JHEP02(2021)194},
	eprint       = {2012.02059},
	archiveprefix = {arXiv},
	primaryclass = {hep-th}
}

@article{Herderschee:2022ntr,
	title        = {{On the differential representation and color-kinematics duality of AdS boundary correlators}},
	author       = {Herderschee, Aidan and Roiban, Radu and Teng, Fei},
	year         = 2022,
	journal      = {JHEP},
	volume       = {05},
	pages        = {026},
	doi          = {10.1007/JHEP05(2022)026},
	eprint       = {2201.05067},
	archiveprefix = {arXiv},
	primaryclass = {hep-th},
	reportnumber = {LCTP-22-01}
}

@article{Armstrong:2022mfr,
	title        = {{New recursion relations for tree-level correlators in anti{\textendash}de Sitter spacetime}},
	author       = {Armstrong, Connor and Gomez, Humberto and Lipinski Jusinskas, Renann and Lipstein, Arthur and Mei, Jiajie},
	year         = 2022,
	journal      = {Phys. Rev. D},
	volume       = 106,
	number       = 12,
	pages        = {L121701},
	doi          = {10.1103/PhysRevD.106.L121701},
	eprint       = {2209.02709},
	archiveprefix = {arXiv},
	primaryclass = {hep-th}
}

@article{Raju:2012zr,
	title        = {{New Recursion Relations and a Flat Space Limit for AdS/CFT Correlators}},
	author       = {Raju, Suvrat},
	year         = 2012,
	journal      = {Phys. Rev. D},
	volume       = 85,
	pages        = 126009,
	doi          = {10.1103/PhysRevD.85.126009},
	eprint       = {1201.6449},
	archiveprefix = {arXiv},
	primaryclass = {hep-th},
	reportnumber = {HRI-ST-1201}
}

@article{Kraus:2006wn,
	title        = {{Lectures on black holes and the AdS(3) / CFT(2) correspondence}},
	author       = {Kraus, Per},
	year         = 2008,
	journal      = {Lect. Notes Phys.},
	volume       = 755,
	pages        = {193--247},
	eprint       = {hep-th/0609074},
	archiveprefix = {arXiv}
}

@article{deBoer:2013gz,
	title        = {{Thermodynamics of higher spin black holes in $AdS_3$}},
	author       = {de Boer, Jan and Jottar, Juan I.},
	year         = 2014,
	journal      = {JHEP},
	volume       = {01},
	pages        = {023},
	doi          = {10.1007/JHEP01(2014)023},
	eprint       = {1302.0816},
	archiveprefix = {arXiv},
	primaryclass = {hep-th}
}

@article{Coussaert:1995zp,
	title        = {{The Asymptotic dynamics of three-dimensional Einstein gravity with a negative cosmological constant}},
	author       = {Coussaert, Oliver and Henneaux, Marc and van Driel, Peter},
	year         = 1995,
	journal      = {Class. Quant. Grav.},
	volume       = 12,
	pages        = {2961--2966},
	doi          = {10.1088/0264-9381/12/12/012},
	eprint       = {gr-qc/9506019},
	archiveprefix = {arXiv},
	reportnumber = {ULB-TH-95-08}
}

@article{Cheung:2017pzi,
	title        = {{TASI Lectures on Scattering Amplitudes}},
	author       = {Cheung, Clifford},
	doi          = {10.1142/9789813233348_0008},
	archiveprefix = {arXiv},
	editor       = {Essig, Rouven and Low, Ian},
	eprint       = {1708.03872},
	primaryclass = {hep-ph},
	reportnumber = {CALT-TH-2017-041}
}

@article{Costa:2011dw,
	title        = {{Spinning Conformal Blocks}},
	author       = {Costa, Miguel and Penedones, Joao and Poland, David and Rychkov, Slava},
	year         = 2011,
	journal      = {JHEP},
	volume       = 11,
	pages        = 154,
	doi          = {10.1007/JHEP11(2011)154},
	archiveprefix = {arXiv},
	eprint       = {1109.6321},
	primaryclass = {hep-th},
	reportnumber = {LPTENS-11-37}
}

@article{Costa:2011mg,
	title        = {{Spinning Conformal Correlators}},
	author       = {Costa, Miguel and Penedones, Joao and Poland, David and Rychkov, Slava},
	year         = 2011,
	journal      = {JHEP},
	volume       = 11,
	pages        = {071},
	doi          = {10.1007/JHEP11(2011)071},
	archiveprefix = {arXiv},
	eprint       = {1107.3554},
	primaryclass = {hep-th},
	reportnumber = {LPTENS-11-22, NSF-KITP-11-128}
}

@book{DiFrancesco:1997nk,
	title        = {{Conformal Field Theory}},
	author       = {Di Francesco, P. and Mathieu, P. and Senechal, D.},
	year         = 1997,
	publisher    = {Springer-Verlag},
	address      = {New York},
	series       = {Graduate Texts in Contemporary Physics},
	doi          = {10.1007/978-1-4612-2256-9},
	isbn         = {978-0-387-94785-3, 978-1-4612-7475-9}
}

@article{Bu:2023cef,
	title        = {{Celestial holography and AdS$_{3}$/CFT$_{2}$ from a scaling reduction of twistor space}},
	author       = {Bu, Wei and Seet, Sean},
	year         = 2023,
	journal      = {JHEP},
	volume       = 12,
	pages        = 168,
	doi          = {10.1007/JHEP12(2023)168},
	eprint       = {2306.11850},
	archiveprefix = {arXiv},
	primaryclass = {hep-th}
}

@article{Bu:2023vjt,
	title        = {{A hidden 2d CFT for self-dual Yang-Mills on the celestial sphere}},
	author       = {Bu, Wei and Seet, Sean},
	year         = 2024,
	journal      = {JHEP},
	volume       = {08},
	pages        = {022},
	doi          = {10.1007/JHEP08(2024)022},
	eprint       = {2310.17457},
	archiveprefix = {arXiv},
	primaryclass = {hep-th}
}

@article{Seet:2025mes,
	title        = {{Single-trace current correlators for 2d models of 4d gluon scattering}},
	author       = {Seet, Sean},
	year         = 2026,
	journal      = {JHEP},
	volume       = {07},
	pages        = 228,
	doi          = {10.1007/JHEP07(2026)228},
	eprint       = {2509.12200},
	archiveprefix = {arXiv},
	primaryclass = {hep-th}
}

@article{He:2015zea,
	title        = {{2D Kac-Moody Symmetry of 4D Yang-Mills Theory}},
	author       = {He, Temple and Mitra, Prahar and Strominger, Andrew},
	year         = 2016,
	journal      = {JHEP},
	volume       = 10,
	pages        = 137,
	doi          = {10.1007/JHEP10(2016)137},
	eprint       = {1503.02663},
	archiveprefix = {arXiv},
	primaryclass = {hep-th}
}

@article{Britto:2004ap,
	title        = {{New Recursion Relations for Tree Amplitudes of Gluons}},
	author       = {Britto, Ruth and Cachazo, Freddy and Feng, Bo},
	year         = 2005,
	journal      = {Nucl. Phys. B},
	volume       = 715,
	pages        = {499--522},
	doi          = {10.1016/j.nuclphysb.2005.02.030},
	archiveprefix = {arXiv},
	eprint       = {hep-th/0412308}
}

@article{Nair:1988bq,
	title        = {{A Current Algebra for Some Gauge Theory Amplitudes}},
	author       = {Nair, V.},
	year         = 1988,
	journal      = {Phys. Lett. B},
	volume       = 214,
	pages        = {215--218},
	doi          = {10.1016/0370-2693(88)91471-2},
	reportnumber = {CU-TP-408}
}

@manual{Rychkov:2016iqz,
	title        = {{EPFL Lectures on Conformal Field Theory in D\ensuremath{>}= 3 Dimensions}},
	author       = {Rychkov, Slava},
	isbn         = {978-3-319-43625-8, 978-3-319-43626-5},
	archiveprefix = {arXiv},
	eprint       = {1601.05000},
	primaryclass = {hep-th},
	reportnumber = {CERN-TH-2016-012}
}

@article{Dirac:1936fq,
	title        = {{Wave equations in conformal space}},
	author       = {Dirac, Paul A. M.},
	year         = 1936,
	journal      = {Annals Math.},
	volume       = 37,
	pages        = {429--442},
	doi          = {10.2307/1968455}
}

@book{Elvang:2015rqa,
	title        = {{Scattering Amplitudes in Gauge Theory and Gravity}},
	author       = {Elvang, Henriette and Huang, Yu-tin},
	year         = 2015,
	publisher    = {Cambridge University Press},
	isbn         = {978-1-316-19142-2, 978-1-107-06925-1}
}

@article{Maldacena:2011nz,
	title        = {{On Graviton Non-Gaussianities During Inflation}},
	author       = {Maldacena, Juan and Pimentel, Guilherme L.},
	year         = 2011,
	journal      = {JHEP},
	volume       = {09},
	pages        = {045},
	doi          = {10.1007/JHEP09(2011)045},
	archiveprefix = {arXiv},
	eprint       = {1104.2846},
	primaryclass = {hep-th},
	reportnumber = {PUPT-2371}
}

@article{CarrilloGonzalez:2022ggn,
	title        = {{Mini-Twistors and the Cotton Double Copy}},
	author       = {Carrillo Gonz\'alez, Mariana and Emond, William and Moynihan, Nathan and Rumbutis, Justinas and White, Chris},
	year         = 2023,
	journal      = {JHEP},
	volume       = {03},
	pages        = 177,
	doi          = {10.1007/JHEP03(2023)177},
	archiveprefix = {arXiv},
	eprint       = {2212.04783},
	primaryclass = {hep-th}
}

@article{Caron-Huot:2021kjy,
	title        = {{Helicity Basis for Three-Dimensional Conformal Field Theory}},
	author       = {Caron-Huot, Simon and Li, Yue-Zhou},
	year         = 2021,
	journal      = {JHEP},
	volume       = {06},
	pages        = {041},
	doi          = {10.1007/JHEP06(2021)041},
	archiveprefix = {arXiv},
	eprint       = {2102.08160},
	primaryclass = {hep-th}
}

@article{Karateev:2017jgd,
	title        = {{Weight Shifting Operators and Conformal Blocks}},
	author       = {Karateev, Denis and Kravchuk, Petr and Simmons-Duffin, David},
	year         = 2018,
	journal      = {JHEP},
	volume       = {02},
	pages        = {081},
	doi          = {10.1007/JHEP02(2018)081},
	eprint       = {1706.07813},
	archiveprefix = {arXiv},
	primaryclass = {hep-th},
	reportnumber = {CALT-TH-2017-031}
}

@article{DelDuca:1999rs,
	title        = {{New Color Decompositions for Gauge Amplitudes at Tree and Loop Level}},
	author       = {Del Duca, Vittorio and Dixon, Lance J. and Maltoni, Fabio},
	year         = 2000,
	journal      = {Nucl. Phys. B},
	volume       = 571,
	pages        = {51--70},
	doi          = {10.1016/S0550-3213(99)00809-3},
	eprint       = {hep-ph/9910563},
	archiveprefix = {arXiv},
	reportnumber = {SLAC-PUB-8294, DFTT-53-99}
}

@article{DelDuca:1999iql,
	title        = {{Factorization of Tree QCD Amplitudes in the High-Energy Limit and in the Collinear Limit}},
	author       = {Del Duca, Vittorio and Frizzo, Alberto and Maltoni, Fabio},
	year         = 2000,
	journal      = {Nucl. Phys. B},
	volume       = 568,
	pages        = {211--262},
	doi          = {10.1016/S0550-3213(99)00657-4},
	eprint       = {hep-ph/9909464},
	archiveprefix = {arXiv},
	reportnumber = {DFTT-45-99}
}

@article{Baumann:2024ttn,
	title        = {{A New Twist on Spinning (A)dS Correlators}},
	author       = {Baumann, Daniel and Mathys, Gr\'egoire and Pimentel, Guilherme L. and Rost, Facundo},
	year         = 2025,
	journal      = {JHEP},
	volume       = {01},
	pages        = 202,
	doi          = {10.1007/JHEP01(2025)202},
	eprint       = {2408.02727},
	archiveprefix = {arXiv},
	primaryclass = {hep-th}
}

@article{Knizhnik:1984nr,
	title        = {{Current Algebra and Wess-Zumino Model in Two-Dimensions}},
	author       = {Knizhnik, V. G. and Zamolodchikov, A. B.},
	year         = 1984,
	journal      = {Nucl. Phys. B},
	volume       = 247,
	pages        = {83--103},
	doi          = {10.1016/0550-3213(84)90374-2},
	editor       = {Khalatnikov, I. M. and Mineev, V. P.}
}

@article{Belavin:1984vu,
	title        = {{Infinite Conformal Symmetry in Two-Dimensional Quantum Field Theory}},
	author       = {Belavin, A. A. and Polyakov, Alexander M. and Zamolodchikov, A. B.},
	year         = 1984,
	journal      = {Nucl. Phys. B},
	volume       = 241,
	pages        = {333--380},
	doi          = {10.1016/0550-3213(84)90052-X},
	editor       = {Khalatnikov, I. M. and Mineev, V. P.},
	reportnumber = {CERN-TH-3827}
}

@article{Bala:2025gmz,
	title        = {{3D Conformal Field Theory in Twistor Space}},
	author       = {Bala, Aswini and Jain, Sachin and S., Dhruva K. and Mazumdar, Deep and Singh, Vibhor},
	year         = 2025,
	month        = 2,
	eprint       = {2502.18562},
	archiveprefix = {arXiv},
	primaryclass = {hep-th}
}

@article{Pasterski:2017ylz,
	title        = {{Gluon Amplitudes as 2d Conformal Correlators}},
	author       = {Pasterski, Sabrina and Shao, Shu-Heng and Strominger, Andrew},
	year         = 2017,
	journal      = {Phys. Rev. D},
	volume       = 96,
	number       = 8,
	pages        = {085006},
	doi          = {10.1103/PhysRevD.96.085006},
	eprint       = {1706.03917},
	archiveprefix = {arXiv},
	primaryclass = {hep-th}
}

@article{De:2026shn,
    author = "De, Shounak and Lee, Hayden",
    title = "{The Vasiliev Grassmannian}",
    eprint = "2603.24656",
    archivePrefix = "arXiv",
    primaryClass = "hep-th",
    month = "3",
    year = "2026"
}

@article{Bala:2026bdx,
    author = "Bala, Aswini and Jain, Sachin and S., Dhruva K. and Rao, Adithya A.",
    title = "{Super-Grassmannians for $\mathcal{N}=2$ to $4$ SCFT$_3$: From AdS$_4$ Correlators to $\mathcal{N}=4$ SYM scattering Amplitudes}",
    eprint = "2604.07503",
    archivePrefix = "arXiv",
    primaryClass = "hep-th",
    month = "4",
    year = "2026"
}

@article{Huang:2026tsh,
    author = "Huang, Yu-tin and Kuo, Chia-Kai and Liu, Yohan and Mei, Jiajie",
    title = "{Beyond Discontinuities: Cosmological WFCs and the Supersymmetric Orthogonal Grassmannian}",
    eprint = "2604.08512",
    archivePrefix = "arXiv",
    primaryClass = "hep-th",
    month = "4",
    year = "2026"
}

@article{Bala:2026trw,
    author = "Bala, Aswini and Jain, Sachin and S, Dhruva K.",
    title = "{The Conformal Grassmannian: A Symplectic Bi-Grassmannian for $CFT_ 4$ Correlators}",
    eprint = "2605.06811",
    archivePrefix = "arXiv",
    primaryClass = "hep-th",
    month = "5",
    year = "2026"
}

@article{Arundine:2026myr,
    author = "Arundine, Mattia and Pimentel, Guilherme L.",
    title = "{Cosmological Collider in the Grassmannian}",
    eprint = "2605.21581",
    archivePrefix = "arXiv",
    primaryClass = "hep-th",
    month = "5",
    year = "2026"
}

@article{Arundine:2026fbr,
    author = "Arundine, Mattia and Baumann, Daniel and Lee, Mang Hei Gordon and Pimentel, Guilherme L. and Rost, Facundo",
    title = "{The Cosmological Grassmannian}",
    eprint = "2602.07117",
    archivePrefix = "arXiv",
    primaryClass = "hep-th",
    month = "2",
    year = "2026"
}

@article{Arundine:2026qqg,
    author = "Arundine, Mattia and Cortes, Veronica Calvo and Koefler, Joris and Rost, Facundo",
    title = "{Positive Geometry of Yang-Mills Correlators}",
    eprint = "2608.27594",
    archivePrefix = "arXiv",
    primaryClass = "hep-th",
    month = "8",
    year = "2026"
}

@article{Bala:2025jbh,
    author = "Bala, Aswini and Jain, Sachin and S., Dhruva K. and Mazumdar, Deep and Singh, Vibhor and Thakkar, Brijesh",
    title = "{A supertwistor formalism for $ \mathcal{N}=1,2,3,4 $ SCFT$_{3}$}",
    eprint = "2503.19970",
    archivePrefix = "arXiv",
    primaryClass = "hep-th",
    doi = "10.1007/JHEP05(2026)288",
    journal = "JHEP",
    volume = "05",
    pages = "288",
    year = "2026"
}

@article{Bala:2025qxr,
    author = "Bala, Aswini and S, Dhruva K.",
    title = "{An Ode to the Penrose and Witten transforms in twistor space for 3D CFT}",
    eprint = "2505.14082",
    archivePrefix = "arXiv",
    primaryClass = "hep-th",
    doi = "10.1007/JHEP11(2025)056",
    journal = "JHEP",
    volume = "11",
    pages = "056",
    year = "2025"
}

@article{Ansari:2025fvi,
    author = "Ansari, Arhum and Jain, Sachin and S, Dhruva K.",
    title = "{AdS$_{4}$ boundary Wightman functions in twistor space: factorization, conformal blocks and a double copy}",
    eprint = "2512.04172",
    archivePrefix = "arXiv",
    primaryClass = "hep-th",
    doi = "10.1007/JHEP06(2026)024",
    journal = "JHEP",
    volume = "06",
    pages = "024",
    year = "2026"
}

@article{Bala:2026hdm,
    author = "Bala, Aswini and Jain, Sachin and S., Dhruva K. and Rao, Adithya A.",
    title = "{The $\mathcal{N}=1$ Super-Grassmannian for CFT$_3$ and a Foray on AdS and Cosmological Correlators}",
    eprint = "2604.07446",
    archivePrefix = "arXiv",
    primaryClass = "hep-th",
    month = "4",
    year = "2026"
}

@article{Bala:2026hag,
    author = "Bala, Aswini and Jain, Sachin and Mazumdar, Deep and Rao, Adithya A.",
    title = "{$\mathbb{R}$eal Ambitwistors {\&} Massive Twistors for CFT$_4$}",
    eprint = "2608.18210",
    archivePrefix = "arXiv",
    primaryClass = "hep-th",
    month = "8",
    year = "2026"
}

@article{CarrilloGonzalez:2026phk,
    author = "Carrillo Gonz{\'a}lez, Mariana",
    title = "{Color-Kinematics Duality and the Double Copy in Curved Spacetimes}",
    eprint = "2608.23686",
    archivePrefix = "arXiv",
    primaryClass = "hep-th",
    reportNumber = "Imperial/TP/2026/MC/04",
    month = "8",
    year = "2026"
}
}

\end{document}